\documentclass{elsart}

\usepackage{psfig}
\usepackage[figuresright]{rotating}

\begin{document}

\begin{frontmatter}

\title{From the Bethe-Salpeter equation to non-relativistic approaches  with  
effective two-body interactions}

\author[n1]{A. Amghar}, 
\author[n2]{B. Desplanques}, \ead{desplanq@isn.in2p3.fr}
\author[n2]{L. Theu{\ss}l} \ead{lukas@isn.in2p3.fr}
\address[n1]{Facult\'e des Sciences, Universit\'e de Boumerdes, 
35000 Boumerdes, Algeria}
\address[n2]{Institut des Sciences Nucl\'eaires\thanksref{m1},   
F-38026 Grenoble Cedex, France}
\thanks[m1]{UMR CNRS/IN2P3-UJF}

\begin{abstract}
It is known that binding energies calculated from the Bethe-Salpeter 
equation in ladder approximation can be reasonably well 
accounted for by an energy-dependent interaction, at least for the lowest 
states. It is also known that none of these approaches gives results 
close to what is obtained by using the same interaction in the so-called 
instantaneous approximation, which is often employed in non-relativistic 
calculations. However, a recently proposed effective interaction was shown to 
account for  the main features of both the Bethe-Salpeter equation and the 
energy-dependent approach. In the present work, a detailed comparison of these 
different methods for calculating binding energies of a two-particle system is 
made. Some improvement, previously incorporated for the zero-mass boson case in 
the derivation of the effective interaction, is also employed for massive 
bosons. The constituent particles are taken to be distinguishable and spinless. 
Different masses of the exchanged boson (including a zero mass) as well as  
states with different angular momenta are considered and the contribution of the 
crossed two-boson exchange diagram is discussed. With this respect, the role 
played by the charge of the exchanged boson is emphasized. It is shown that the 
main difference between the Bethe-Salpeter results and the instantaneous 
approximation ones are not due to relativity as often conjectured.
\end{abstract}

\end{frontmatter}

\section{Introduction}
Understanding the discrepancies for binding energies obtained from 
different approaches, ranging from the ladder Bethe-Salpeter equation at one 
extreme \cite{BETH} to a non-relativistic equation with a so-called 
instantaneous one-boson exchange interaction at the other extreme, has been and 
is still a subject which raises considerable interest and curiosity 
\cite{SILV,BILA,PHIL,NIEU,FRED,BIJT,DARE,AMGH,THEU}. Although it is more founded 
theoretically, the 
Bethe-Salpeter equation in its simplest form (along with other approaches) is 
not doing as well as the Dirac or Klein-Gordon equation in reproducing the 
spectrum of atomic systems. These last equations assume a potential which is 
apparently consistent with an instantaneous propagation of the exchanged boson 
and their success has certainly contributed much in founding this approximation 
as a model for non-relativistic interactions.

It has been suggested that the problem was specific to QED together with the 
one-body limit \cite{FELD} and irrelevant outside this case, when bosons are 
massive or the one-body limit does not apply. This is also supported by a number 
of works that motivated the instantaneous approximation from the 
Bethe-Salpeter equation. Although proofs are not transparent and rely on some 
approximations, allowing authors to have doubt \cite{DARE}, it is 
conjectured that including the contribution of crossed-boson exchange terms
would resolve the problem \cite{TODO,FRIA}. In this context, a few benchmark 
results emerge. It was clearly stated  that the Bethe-Salpeter equation applied 
to the Wick-Cutkosky model \cite{WICK} does not support the validity of the 
instantaneous approximation \cite{SILV}. Instead, there is agreement with 
results from an energy-dependent interaction \cite{BILA,PHIL}. The role of 
crossed boson exchange, as estimated using the Feynman-Schwinger representation 
(FSR) method \cite{NIEU}, is also important in the case of massive bosons with 
equal-mass constituents.

Quite recently, two of the present authors, starting from a field-theory based 
approach, derived an effective interaction to be used in a non-relativistic 
framework \cite{AMGH}. It was found that this interaction, also obtained on a 
different basis in earlier works \cite{FST}, involved a renormalization of the 
so-called instantaneous approximation. For small couplings, the renormalization 
factor is given by the probability of the system being in a two-body component, 
without accompanying bosons in flight. A close relation between wave functions 
issued from the field-theory (energy-dependent) scheme and the non-relativistic 
scheme (using an effective interaction) was found. It was suggested that the 
spectra obtained from the Bethe-Salpeter or other field-theory based approaches 
and that one resulting from the effective interaction could be quite similar for 
the lowest states. The role of crossed diagrams was also considered. In the case 
of a neutral, spinless boson it was found that their contribution largely 
cancels the renormalization of the instantaneous approximation 
interaction. While this result apparently provided support for the instantaneous 
approximation, its specific character was demonstrated by considering a model 
involving the exchange of ``charged'' bosons, which evidences a totally 
different pattern. Here, the contribution of crossed-boson exchange not only 
compensated the effect of the renormalization but provided even a lot more 
attraction in some cases. A first confirmation using the  Bethe-Salpeter 
equation was obtained \cite{THEU}.

In this work, we consider a system of two scalar, distinguishable constituents 
that interact by the exchange of a scalar boson. This is the simplest case one 
can think of. It allows one to minimize complications with spin, like Z-type 
contributions for instance, which will obscure conclusions. We essentially want 
to compare binding energies with the main intent to establish a close 
relationship between the different approaches. This completes the work performed 
in ref. \cite{AMGH} for wave functions and extends it to the Bethe-Salpeter 
equation \cite{THEU}.  We will in particular study how this relationship is 
fulfilled when going from the  ladder approximation to a more complete case 
involving crossed-boson exchange contributions. Regarding the latter, 
we will have in mind as benchmarks the results of Nieuwenhuis and Tjon 
\cite{NIEU} as well as those obtained in the so-called instantaneous one-boson 
exchange approximation. Our study will concentrate on systems with a rather 
small binding energy, of the order of what is reasonably accessible for a 
non-relativistic approach. An important issue is to determine which physics is 
behind some equation, with the idea that information obtained from the 
Bethe-Salpeter  equation and a non-relativistic approach, the one more founded 
theoretically, the other closer to a physical interpretation, can complement 
each other. Quite generally, our aim is to derive effective interactions in a 
given scheme. This motivation has some similarity with the one underlying the 
work of Bhaduri and Brack who tried to reproduce the spectrum of the Dirac 
equation with an effective interaction in a non-relativistic approach 
\cite{BHAD}. The difference between this work and ours is in the dynamics: 
coupling of large and small components in one case, coupling of two-body 
components with components involving mesons in flight in the other. 
A renormalization of the instantaneous one-boson exchange interaction has been 
considered by Lepage \cite{LEPA}. This work has in common with the present one 
that it 
takes into account physics beyond some scale but it differs in the way how 
this physics 
is incorporated, parametrized in one case, theoretically motivated in the other.

The plan of the paper is  as follows. In the second part, we remind essential
theoretical ingredients relative to the different approaches that are referred 
to later on. These ones include the Bethe-Salpeter equation, an energy-dependent 
interaction resulting from time-ordered diagrams, an approach using an 
effective energy-independent interaction and finally an interaction in the
instantaneous approximation. In the third section we present
results obtained in the ladder approximation,
incorporating higher-order box contributions in the effective approaches
to allow for a consistent comparison with the Bethe-Salpeter results.
Cases with massive as well as zero-mass bosons (Wick-Cutkosky model) 
are considered. The contribution due to crossed-boson exchange is  
examined and discussed in the fourth section. Two cases, corresponding to the 
exchange of neutral and charged bosons, are studied. The fifth section finally 
contains a general discussion and conclusions.

\section{Descriptions of two-body systems: different approaches}
In this section, we  present the various approaches that will be used 
in the next sections to calculate binding energies of two-body systems. 
The first one is based on the Bethe-Salpeter equation which is well 
known for its manifestly  relativistic covariant character. The field-theory 
foundation is more transpa\-rent in the second approach which is based on 
time-ordered diagrams and, therefore, gives an interaction that depends 
on the total energy of the system, $E$. This energy dependence is 
a reminiscent of the fact that the two-body system under 
consideration is coupled to components with bosons in flight. 
The third approach, which is derived from the previous one,
is based on an effective interaction where the energy dependence has been
eliminated. It differs in particular from the instantaneous one-boson exchange 
approximation, which is frequently employed in non-relativistic descriptions 
of two-body systems. 

\subsection{Bethe-Salpeter equation}

The expression of the Bethe Salpeter equation for a system of scalar particles 
of mass $m$, interacting by the exchange of bosons of mass $\mu$, is well known.
For definiteness, we nevertheless remind it here. It is given in momentum space 
by:
\begin{eqnarray}
\left(m^2-(\frac{P}{2}+p)^2\right)\,\left(m^2-(\frac{P}{2}-p)^2\right) \; 
\Phi(p,P) =  \hspace{2cm} \nonumber  \\
    i \int \frac{d^4p'}{(2\pi)^4} \, K(p,p',P) \, \Phi(p',P).
   \label{Aa}
\end{eqnarray}
In this equation, $P$, $p$ and $p'$ respectively denote the total momentum of 
the system and half of the relative momenta.  
\begin{figure}[ht]
\begin{center}
\mbox{\psfig{file=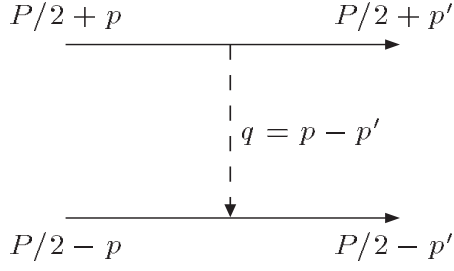,width=6cm,height=3.5cm}}  
\end{center}
\caption{Graphical representation of the single-boson exchange contribution to 
the two-body interaction.\label{fig1}}
\end{figure}
$K(p,p',P)$  represents the interaction kernel. To lowest order it contains a 
well determined contribution due to one-boson exchange (see Fig. \ref{fig1} for 
a graphical representation), its expression being given by:
\begin{equation}
 K^{(1)} = -\frac{\tilde{g}^{2} \; O_{1} \, O_{2} }{  \mu^{2} - t} 
 \equiv -\frac{\tilde{g}^{2} \; O_{1} \, O_{2} }{  \mu^{2} - (p-p')^2 
-i\epsilon}.
 \label{Ab}
\end{equation}
Restricting the full kernel to this lowest order term, one obtains the 
ladder approximation of the Bethe-Salpeter equation, which has been extensively 
used in the literature. The coupling, $\tilde{g}^2$ in Eq. (\ref{Ab}), has the 
dimension of a mass squared. The quantity $\frac{\tilde{g}^2}{4m^2}$ is directly 
comparable to the coupling constant $g^2$ which will be used in the following, 
to the coupling constant $g^2_{MNN}$,  frequently used in hadronic physics, or 
to the quantity $4\pi\, \alpha$, where $\alpha$ is the usual QED coupling 
constant.  Anticipating the following, we introduced the operators $ O_{1} \, 
O_{2} $ in Eq. (\ref{Ab}), that may be equal to $1_1 \, 1_2 $ or 
$\vec{\tau}_1 \cdot \vec{\tau}_2$ in the present work. The first case 
corresponds to the exchange of a neutral boson and the factor may 
be simply omitted. The second case corresponds to the exchange of bosons 
carrying some kind of isospin, equal to 1, while the constituents would have 
isospin $\frac{1}{2}$. The complete interaction kernel employed in Eq. 
(\ref{Aa}) also contains multi-boson exchange contributions that Eq. (\ref{Ab}) 
cannot account for, namely of the non-ladder type. As the simplest example of 
such contributions, we shall consider in Sect. 4 the crossed two-boson exchange. 
The numerical method we used to solve Eq. (\ref{Aa}) and a detailed discussion 
was given in ref. \cite{THEU}.

\subsection{Energy-dependent interaction from field-theory based 
one-boson exchange}

\begin{figure}[ht]
\begin{center}
\mbox{\psfig{file=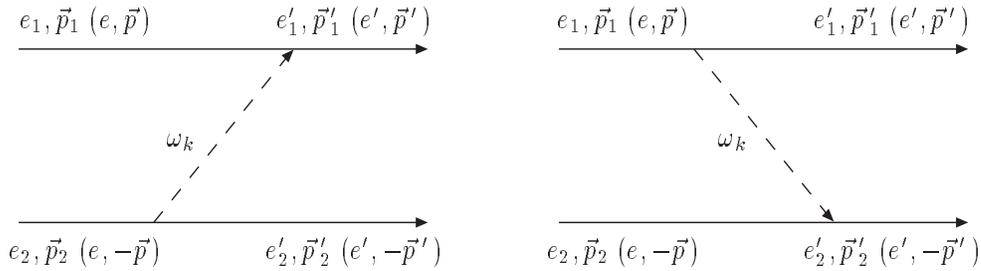,width=13cm,height=3.5cm}}  
\end{center}
\caption{Time-ordered single-boson exchange contributions to the two-body 
interaction with indication of the kinematics in an arbitrary frame and in the 
center of mass (between parentheses).\label{fig2}}
\end{figure}

The single-boson exchange contribution we want to consider is represented by 
the time-ordered diagrams of Fig. \ref{fig2} . 

The corresponding interaction is easily 
obtained from second-order perturbation theory:
\begin{eqnarray}
V^{(1)}_E(,\vec{p}_2,\vec{p}\,'_1,\vec{p}\,'_2)= \hspace{8cm} 
\nonumber \\
\frac{g^{2} \; O_{1} \, O_{2} \,}{2 \, \omega_{k}} \, 
\sqrt{ \frac{m^2}{ e_1 e_2} } \, 
\left( \frac{1}{E-\omega_{k}-e_1-e'_2}
 +     \frac{1}{E-\omega_{k}-e_2-e'_1} \right) \, 
\sqrt{ \frac{m^2}{ e'_1 e'_2 } },
\label{Ba}
\end{eqnarray}
with $\omega_{k} = \sqrt{\mu^{2}+(\vec{p}_1-\vec{p}\,'_1)^{\,2}},\;
e_i= \sqrt{m^{2}+\vec{p_i}^{\,2}} $ and $  e_i'= \sqrt{m^{2}+\vec{p_i}'^{\,2}}.$ 
When the energy is conserved, this equation allows one to recover the standard 
Feynman propagator. In the center of mass, the two terms representing the 
contributions of the diagrams displayed in Fig. \ref{fig2} are equal. Equation 
(\ref{Ba}) then simplifies and reads:
\begin{equation}
V^{(1)}_E(\vec{p},\vec{p}\,')= 
\frac{g^{2} \; O_{1} \, O_{2} \,}{\omega_{k}} \, \frac{m}{e} \, 
\left( \frac{1}{E-\omega_{k}-e-e'}  \right) \, \frac{m}{e'},
\label{Bb}
\end{equation}
with $\omega_{k} = \sqrt{\mu^{2}+\vec{k}^{\,2}},\; 
\vec{k}=\vec{p}-\vec{p}\,',\;
e= \sqrt{m^{2}+\vec{p}^{\,2}} $ and $
e'= \sqrt{m^{2}+\vec{p}\,'^{\,2}}, \; $
where $\vec{p}$ and $\vec{p}\,'$ represent the relative 3-momenta of the 
constituent
particles. The two normalization factors, $\frac{m}{e}$ and $\frac{m}{e'}$ in 
Eq. (\ref{Bb}), are appropriate for the case where the total free energy of the 
constituent particles has the semi-relativistic expression, 
$e=\sqrt{m^{2}+\vec{p}^{\,2}}$, 
the corresponding equation to be fulfilled being of the Klein type \cite{KLEI}:
\begin{equation}
(2 \,e - E) \, \psi(\vec{p}\,) =
-\int \frac{d\vec{p}\,'}{(2\pi)^3} \; 
\left(V^{(1)}_E(\vec{p},\vec{p}\,')+ \dots \right) \, \psi(\vec{p}\,').
\label{Bc}
\end{equation}
The dots represent multi-boson exchange terms, some of which  will be 
expli\-citly considered in the following. An important feature of the 
interaction (\ref{Bb}) is its dependence on the total energy of the system, $E$, 
which prevents one from using it in a Schr\"odinger equation without some 
caution. This energy dependence has the same origin as the one in the full Bonn 
model of the nucleon-nucleon interaction \cite{BONN}. It is reminded that the 
above equation may be obtained from the Bethe-Salpeter equation, Eq. (\ref{Aa}), 
by putting: 
\begin{eqnarray}
\left(m^2-(\frac{P}{2}+p)^2\right)\,\left(m^2-(\frac{P}{2}-p)^2\right) \; 
\Phi(p,P) 
\hspace{4cm}\nonumber \\
\simeq \left( 2 \, e \right)^2 \; 
\left( e-\frac{E}{2}-p_0 \right) \;
 \left( e-\frac{E}{2}+p_0 \right) \; \Phi(p,P) \hspace{1cm}
\nonumber \\
 \equiv \frac{2\,e}{E} \; (2e-E)\,\psi(\vec{p}\,), \hspace{1cm}
 \label{Bd}
\end{eqnarray}
consistently neglecting the dependence on $p_0$ (but not $p'_0$!) in the kernel 
as well as corrections of order $\frac{E-2e}{2\, E}$ (we assume that the mass of 
the exchanged boson, $\mu$, is smaller than the constituent mass, $m$). In 
accordance with the center of mass validity of Eq. (\ref{Bb}), $P_0$ has been 
replaced by $E$. It is also reminded that an equation like Eq. (\ref{Bc}) (or 
similar ones used below) can always be cast into the form of a 
4-dimensional, but non-covariant equation, as it can be checked from the 
following:
\begin{equation}
(2e-E) \; \psi(\vec{p}\,) =
i\int \frac{d^4 p \,'}{(2\pi)^4} \;\,  
\frac{\left(V^{(1)}_E(\vec{p},\vec{p}\,')+ \dots \right) \;\;(2e'-E)\, 
\psi(\vec{p}\,')}{\left( 
e'-\frac{E}{2}-p'_0+i\epsilon \right) \;
 \left( e'-\frac{E}{2}+p'_0+i\epsilon \right) }.
\label{Bdd}
\end{equation}

A non-relativistic, but still energy-dependent form of Eq. (\ref{Bc}) may be 
used:
\begin{equation}
\left(\frac{p^2}{m} - (E-2m)\right) \, \psi(\vec{p}\,) =
-\int \frac{d\vec{p}\,'}{(2\pi)^3} \; 
\left(V^{(1)}_E(\vec{p},\vec{p}\,')+ \dots \right) \; \psi(\vec{p}\,'),
\label{Be}
\end{equation}
where the normalization factors $\frac{m}{e}$ appearing in  $V^{(1)}_E$ should 
be consistently neglected. 
This last equation represents an intermediate step in our 
effort to derive an effective non-relativistic interaction in the next 
subsection.

\begin{figure}[htb]
\begin{center}
\mbox{\psfig{file=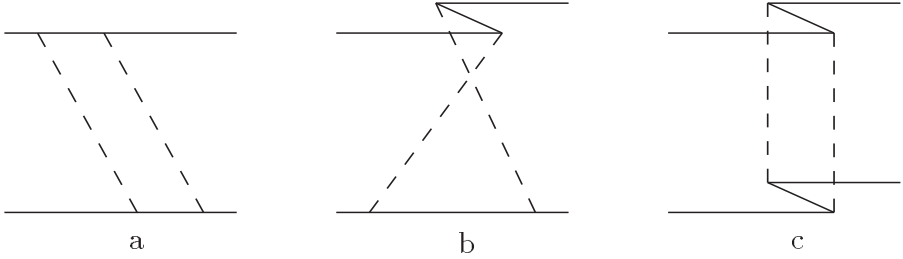,width=12cm,height=3.5cm}}  
\end{center}
\caption{Box type two-boson exchange contributions to the two-body 
interaction in time-ordered perturbation theory.\label{fig3}}
\end{figure}

In order to make a meaningful comparison with the ladder Bethe-Salpeter 
approach, box-type multi-boson exchange contributions to the interaction $V_E$ 
should be considered. Some of them are shown in Fig. \ref{fig3}. The most 
important one corresponds to diagram a) in this figure. Its contribution is of 
order $\left(\frac{1}{m}\right)^0$ and has the same magnitude as the off-shell 
effects due to the presence of the term $E-e-e'$ in the denominator of Eq. 
(\ref{Bb})  (see also ref. \cite{AMGH}). Its expression is given by:
\begin{eqnarray}
\nonumber
V^{(2a)}(\vec{p\,},\vec{p}\,') =- \frac{g^4}{2} \int \frac{d 
\vec{q}}{(2\pi)^3} \;   \frac{m}{e} \frac{(O_{1} \, O_{2}) \, 
(O_{1} \, O_{2})}{\omega \, \omega'} \frac{m}{e'} \left(\frac{m}{e_q}\right)^2 
\hspace{2cm} \\ \times \frac{1}{\left(\omega+e+e_q-E\right)} 
\frac{1}{\left(\omega'+e'+e_q-E\right)} 
\frac{1}{\left(\omega+\omega'+e+e'-E\right)},
 \label{Bf} 
\end{eqnarray}
with  
$\omega=\sqrt{\mu^{2}+(\vec{p}-\vec{q}\,)^{2}}$, 
$\omega'=\sqrt{\mu^{2}+(\vec{p}\,'-\vec{q}\,)^{2}}$  and 
$e_q=\sqrt{m^{2}+\vec{q}\,^{\,2}}$. The quantity $\vec{q}$ represents the 
momentum of the constituent particle in the intermediate state and 
the operators $O_{i}$ are put between parentheses to remind that they do not 
necessarily commute. As we do not pay much attention to contributions of order 
$g^6$, one could be tempted to drop terms like  $E-e-e'$ in the denominator of 
Eq. (\ref{Bf}), which are of the same order. Doing this also has the advantage 
that the integral of Eq. (\ref{Bf}) can be worked out analytically (see Appendix 
\ref{appA}). We shall nevertheless retain the full expression of Eq. (\ref{Bf}) 
in our calculations, in order to deal consistently with the energy dependence
of the box diagram. Numerically, the effect of the latter one turned out to be
substantial. On the other hand, contributions of Z-diagrams like Fig. \ref{fig3} 
b,c have a typical relativistic character and are of order 
$\left(\frac{1}{m}\right)^1$ and $\left(\frac{1}{m}\right)^3$, respectively. 

The expression of the interaction given by Eq. (\ref{Bb}) is much simpler than 
the one used in ref. \cite{BILA}, which was however derived on a 
different basis. The latter one most probably incorporates higher order 
corrections like 
those of Fig. \ref{fig3} b,c, and, furthermore, should be used with a different 
equation.

\subsection{Energy-independent approach and effective interaction}  
In order to get an energy-independent interaction from the general interaction 
kernel given by Eq. (\ref{Bb}), the energy dependence of the latter one 
was appro\-ximated in ref. \cite{AMGH} by a linear term (see ref. \cite{DESP} 
for examples in relation with the present work):

\begin{equation}
\frac{1}{  E-2m -\omega_{k} -\frac{\vec{p}^{\,2}}{2m} 
-\frac{\vec{p}\,'^{\,2}}{2m}  }=
-\frac{1}{\omega_{k}}
-\frac{ E-2m-\frac{\vec{p}^{\,2}}{2m}-\frac{\vec{p}\,'^{\,2}}{2m} }{ 
\omega^{2}_{k} 
}+\cdots ,
\label{Ca}
\end{equation}
where we replaced, as we shall do from now on, the relativistic 
expressions for the kinetic energies by their non-relativistic counterparts.
The second term on the right hand side of Eq. (\ref{Ca}), 
after insertion in Eqs. (\ref{Bc}) or (\ref{Be}), was then removed by 
a transformation which is formally similar to the one introduced by Perey and 
Saxon \cite{PERE} in order to change a linear energy-dependent term of an 
optical potential into a dependence on the squared momentum and vice versa. The 
inverse transformation is used for solving an equation with a squared 
momentum-dependent term when employing the Numerov algorithm. 
In both cases, it has no definite physical ground beyond its mathematical 
character. The transformation which was performed here is more in the spirit of 
the Foldy-Wouthuysen one (see ref. \cite{AMGH} for details). It implicitly 
corresponds to introducing effective degrees of freedom with the drawback 
that the interaction generally acquires a non-local character. An expansion 
similar to that given in Eq. (\ref{Ca}) was used by Lahiff and Afnan 
\cite{AFNA}, so that some of our results obtained with the full energy 
dependence of the interaction kernel have  a relationship to theirs. However,  
these authors have not considered the possibility of removing the energy 
dependence by the transformation mentioned above. The resulting 
energy-independent interaction is the main object of studies in the 
present work.

Since the expansion parameter in  Eq. (\ref{Ca}) is roughly equivalent to the 
interaction strength, this equation makes sense as long as the coupling is not 
too large. In the present work, we will improve upon this approximation, 
extending to massive bosons what was done in ref. \cite{AMGH} for the zero-mass 
case. In the latter one, the Coulomb-like potential kept its analytical form, 
only the coupling constant was renormalized by a factor that was determined by 
solving a self-consistency equation. The zero-mass case will be reminded in 
Sect. 
3.2 below, together with further developments.

Starting from the formal expression of the interaction in configuration space:
\begin{equation}
V_E(r) = g^{2} \; O_{1} \, O_{2} \; \frac{1}{2\pi^{2}}
  \int \frac{dk k^{2} j_{0}(kr)}{\omega_{k}\left( E-2m -\omega_{k} 
-\frac{\vec{p}^{\,2}}{2m} -\frac{\vec{p}\,'^{\,2}}{2m} \right)},
  \label{Cz}
\end{equation}
we assume that the quantity $E-2m  -\frac{\vec{p}^{\,2}}{2m} 
-\frac{\vec{p}\,'^{\,2}}{2m}$ in the denominator under the integral can be 
approximated by the effective potential we want to determine. This procedure is 
often used in other domains of physics where it underlies for instance 
the Thomas-Fermi-, WKB-, or eikonal approximations \cite{RING,MAHA}, and  it
immediately provides a self-consistency equation:
\begin{equation}
V^{nr}_{eff,sc}(r)=-g^{2} \; O_{1} \, O_{2} \; \frac{1}{2\pi^{2}}
  \int \frac{dk k^{2} j_{0}(kr)}{\omega_{k}
  \left(\omega_{k} -V^{nr}_{eff,sc}(r)\right)}.
  \label{Cy}
\end{equation}
By neglecting the effective interaction under the integral, 
one recovers the standard expression of the instantaneous 
one-boson exchange potential:
\begin{equation}
V_{0}(r)=-g^{2} \, O_{1} \, O_{2} \;     \frac{ e^{-\mu r} }{4\pi r}.
\label{Cd}
\end{equation}
One can go a step further and make an expansion of the r.h.s. of Eq. (\ref{Cy}) 
up to first order in the effective potential:
\begin{equation}
V^{nr}_{eff}(r)=V_{0}(r)-V^{nr}_{eff}(r)\;V_{1}(r),
  \label{Cx}
\end{equation}
where 
\begin{equation}
V_{1}(r) = g^{2} \; O_{1} \, O_{2} \; \frac{1}{2\pi^{2}}
  \int \frac{dk k^{2} j_{0}(kr)}{(\omega_{k})^{3}}
  =  \frac{g^{2} \; O_{1} \, O_{2}}{2\pi^{2}}\; K_0(\mu r).
  \label{Ce}
\end{equation}
By rearranging terms in Eq. (\ref{Cx}), one recovers the expression of the 
effective interaction given in ref. \cite{AMGH}:
\begin{equation}
V^{nr}_{eff}(r)=\frac{V_{0}(r)+...}{\left(1+V_{1}(r)+...\right)},
\label{Ci}
\end{equation}
where the dots may account for two, three, ... boson exchange. This expression 
has a simple interpretation in the limit of small couplings. It corresponds to 
the instantaneous approximation potential renormalized by the probability that 
the system is in a two-body component. Notice that the self-consistent potential 
of Eq. (\ref{Cy}) can be put in a form analogous to Eq. (\ref{Ci}), with a  
generalized renormalization factor:
\begin{equation}
V^{nr}_{eff,sc}(r)=\frac{V_{0}(r)+...}{1+ 
\frac{g^{2} \; O_{1} \, O_{2}}{2\pi^{2}}
  \int \frac{dk k^{2} 
j_{0}(kr)}{\omega_{k}^{2}\left(\omega_{k}-V^{nr}_{eff,sc}(r)\right)}+...}.
\label{Cj}
\end{equation}

While making the above developments, we assumed that the operator $p^2$ could be 
treated as a number. To go beyond this approximation, an expansion of the 
denominator in Eq. (\ref{Cz}) should be performed in the vicinity of
\begin{equation} 
E-2m  -\frac{\vec{p}^{\,2}}{2m} -\frac{\vec{p}\,'^{\,2}}{2m} \simeq 
V^{nr}_{eff,sc}(r).
\label{Cv}
\end{equation}
This approximate equality is valid provided it is 
applied to a solution of a Schr\"odinger equation. 
One thus gets up to first order in the expansion:
\begin{equation}
V_E(r) = V^{nr}_{eff,sc}(r)+ \frac{1}{2} \left\{   
\left(\frac{\vec{p}^{\,2}}{m}+V^{nr}_{eff,sc}(r)+2m-E\right), 
V_{1,sc}(r)\right\},
  \label{Ct}
\end{equation}
where 
\begin{equation}
V_{1,sc}(r)=g^{2} \; O_{1} \, O_{2} \; \frac{1}{2\pi^{2}}
  \int \frac{dk k^{2} 
j_{0}(kr)}{\omega_{k}\left(\omega_{k}-V^{nr}_{eff,sc}(r)\right)^{2}}
  \label{Cu}
\end{equation}
generalizes the result given by Eq. (\ref{Ce}). The expression of Eq. (\ref{Ct}) 
can now be inserted in the equation that corresponds to Eq. (\ref{Be}) in 
configuration space. One thus gets the following equation with an 
energy-dependent interaction:
\begin{eqnarray}
\Bigg( V^{nr}_{eff,sc}(r) ( 1+V_{1,sc}(r) ) + \frac{1}{2} 
\left\{ \frac{\vec{p}^{\,2}}{m},\Big(1+V_{1,sc}(r) \Big) \right\} \hspace{3cm} 
 \nonumber \\  - (E-2m) \Big(1+V_{1,sc}(r)\Big) \Bigg) \psi(r)=0.
\label{Cb}
\end{eqnarray}
The equation relative to the energy-independent scheme is obtained by 
multiplying Eq. (\ref{Cb}) from the left by $\Big(1+V_{1,sc}(r)\Big)^{-1/2}$ 
and making the substitution:
\begin{equation}
\psi(r)=\Big(1+V_{1,sc}(r)\Big)^{-1/2} \, \phi(r). 
\label{Cf}
\end{equation}
One thus gets a usual non-relativistic Schr\"odinger equation:
\begin{equation}
\left( V_{eff}(r)+\frac{\vec{p}^{\,2}}{m} -(E-2m) \right) \phi(r)=0,
\label{Cg}
\end{equation}
with an effective potential
\begin{equation}
V_{eff}(r)=V^{nr}_{eff,sc}(r)
   -\frac{1}{4m} \frac{\left(\vec{\nabla}V_{1,sc}(r)\right)^2}
   {\Big(1+V_{1,sc}(r)\Big)^2}.
   \label{Ch}
\end{equation}
The (presumably) dominant first term, of order $\left(\frac{1}{m}\right)^0$, 
has a typical non-relativistic character. It can be checked that its 
contribution to order $g^4$ (i.e. including the box diagram, Fig. \ref{fig3} a)  
is identical to the one obtained by substracting from the Feynman two-boson 
exchange diagram that part resulting from iterating the one-boson exchange. This 
represents another way of calculating the contributions to energy-independent 
potentials with the range of two-boson exchange \cite{DURS,PARI}. However, the 
present approach allows one to take into account higher order corrections that 
are relevant when $V_{1,sc}(r)$ becomes large. The second term in 
Eq. (\ref{Ch}) is required for consistency with Eq. (\ref{Cb}) and, most 
importantly, it represents the first correction to the approximation 
given by Eq. (\ref{Cv}) at the denominator of Eq. (\ref{Cz}). Being of a 
higher order in $\frac{1}{m}$, it will be neglected in most calculations using 
the effective interaction. This is consistent with neglecting other
relativistic corrections of the same order, especially the normalization factors 
$\frac{m}{e}$ present in Eq. (\ref{Bb}). These ones provide some non-locality 
that, most probably, could be accounted for in the above developments, but at 
the price of a tremendous numerical work. For the same reason, we will retain, 
for the numerical calculations presented below, the expression of 
the self-consistent, effective potential given by the solution of Eq. (\ref{Cy}). 
When accounting for the higher-order terms in $\frac{1}{m}$, the derivation of 
the effective potential proceeds analogously, by incorporating the second term 
of Eq. (\ref{Ch}) in the self-consistency equation.

The dressed (effective) character of the object described by $\phi(r)$, Eq. 
(\ref{Cg}), is better seen by looking at the expression of the norm in the local 
approximation:
\begin{eqnarray}
N = \int d\vec{r} \; \phi^2(r)
 = \int  d\vec{r} \; \bigg( \frac{\phi^2(r)}{ \Big(1+V_{1,sc}(r)\Big)}+ 
\frac{\phi^2(r)\;V_{1,sc}(r)}{ \Big(1+V_{1,sc}(r)\Big)} \bigg)
 = \hspace{1cm} \nonumber \\ \int d\vec{r} \; \left( \psi^2(r) +  
\psi(r)\,V_{1,sc}(r)\,\psi(r) \right) .
 \label{Cjj}
\end{eqnarray}
The quantity $\phi^2(r)$ describes a hybrid state that is a coherent 
superposition of two components, with weights respectively given by 
$\frac{1}{1+V_{1,sc}(r)}$ and  $\frac{V_{1,sc}(r)}{1+V_{1,sc}(r)}$. 
The first one, which from Eq. (\ref{Cf}) is seen to be described by $\psi(r)$, 
corresponds to the bare two-body component. The second one, which corresponds to 
a three-body component, has a more complicated structure. The relation in 
this case can be obtained from the expression of the component with one boson 
in flight:
\begin{equation}
|{\rm boson}+2 \, {\rm constituents}\rangle= \frac{1}{E-\omega_k-e_1-e_2}\, 
H_I \, |2 \, {\rm constituents}\rangle,
\label{Ck}
\end{equation}
where $H_I$ is the boson-constituent interaction. Using this expression,  
it can be checked that the part corresponding to extra bosons in flight 
(discarding the boson cloud of the constituents) provides a 
contribution to the normalization given by  $\int 
d\vec{r} \; \psi^2(r) (-\frac{\partial V_E(r)}{\partial E})$. 
The identification with the last term in Eq. (\ref{Cjj}) is achieved via the 
expression given by Eq. (\ref{Ct}), or equivalently from an explicit 
calculation starting with Eq. (\ref{Cz}), allowing one to directly obtain Eq. 
(\ref{Cu}).

\subsection{Instantaneous one-boson exchange approximation}  
It is noticed that the potential of Eq. (\ref{Ch}) does not correspond 
to the so-called instantaneous  approximation given by $V_{0}(r)$ in Eq. 
(\ref{Cd}). This one, which is often used 
as  a reference, underlies many models dealing with hadronic systems and, at 
first sight, the well known Coulomb potential. It can be partly justified as 
follows. Starting from the usual covariant expression for a boson propagator in 
configuration space,
\begin{equation}
\Delta(x_1-x_2)= i \int \, \frac{d^4q}{(2\pi)^4} \, 
\frac{e^{iq \cdot (x_1-x_2)}}{\mu^2+\vec{q}^{\,2}-q_0^2-i\epsilon},
\label{Da}
\end{equation}
and assuming that the energy transfer $q_0$ is small, one gets
\begin{equation}
\Delta_B(x_1-x_2) \simeq i \, \delta (t_1-t_2) \, \frac{e^{-\mu \, r}}{4 \pi 
r}, 
\;\;\;({\rm with} \; r=|\vec{r}_1-\vec{r}_2|).
\label{Db}
\end{equation}
The assumption that $q_0$ is small is valid for a Born amplitude in the 
non-relativistic domain. Beyond this approximation, $q_0$, which is  
an integration variable at the r.h.s. of Eq. (\ref{Da}), is not limited 
to small values. It may be large
and, in particular, can give rise to singularities when $q_0= \pm \sqrt{\mu^2 + 
\vec{q}^{\,2}}$. These poles are responsible for the renormalization of 
the so-called instantaneous potential, Eq. (\ref{Cd}), leading to 
the effective potential of Eq. (\ref{Ch}). Their contributions are omitted in 
the spectator approach \cite{GROS} where they could allow one to symmetrize the 
role of the two constituent particles. This omission is generally justified by 
the fact that these extra contributions are cancelled by 
crossed-boson exchange diagrams, thus recovering an instantaneous-like 
single-boson exchange interaction. However, this only works for neutral, 
spinless bosons, which discards two important cases in hadronic physics:  pions 
and gluons.

While the potentials of Eqs. (\ref{Ch}) and (\ref{Cd}) both have an 
instantaneous character, there is not necessarily a contradiction as they do not 
correspond to the same mathematical description. In one case, one works with 
asymptotic, bare particles that exchange bosons. In the other, one works with 
dressed particles that are eigenstates of the Hamiltonian, incorporating 
components with bosons in flight.  The concept of an instantaneous propagation 
of the interaction is therefore partly scheme dependent. It will become clear 
from our results that the effects implicitly incorporated in the effective 
interaction account at least partially for the dynamics of the original relative 
time (or energy) variable, which has been integrated out (as for instance in Eq. 
(\ref{Bc})). The effects under discussion  may therefore not  
be relativistic ones as sometimes  advocated in the literature. Instead, they 
are quite similar to those appearing in the many-body problem. In the case of 
the nucleon-nucleus interaction for instance, they originate from the energy 
dependence of the optical potential due to the coupling to other channels, whose 
effects can be partly incorporated in mean-field approximations through a 
contribution to the effective mass \cite{MAHA}. The close relationship between 
the quasi-particle strength introduced in that work, $Z=(1-\frac{d}{dE}\Delta 
V(E))^{-1}$, and the quantity $(1+V_{1})^{-1}$ appearing in many equations of 
the present section, is evident.

The non-relativistic character of the effects we are looking at is further 
supported by the examination of the energy denominator obtained from 
time-ordered diagrams, which describes the evolution of the system from 
time $t_1$ to time $t_2$. This quantity contains the energy of the exchanged 
boson $\omega_k$ and, also, the difference of the total energy of the system 
and the energies of the constituents, $E-e-e'$. The non-relativistic limit of 
this last factor can be taken but, appearing on the same footing as the 
$\omega_k$ term, it cannot be put to zero as the so-called instantaneous 
approximation would suppose. At best, it can be considered as a recoil 
correction of order $\frac{\mu}{m}$. This can be seen by expanding the energy 
denominator appearing in Eq. (\ref{Bb}),
\begin{equation}
 E-e-e'-\omega_k= -\mu-
\frac{(\vec{p}-\vec{p}\,')^2}{2\mu}-\frac{\vec{p}^{\,2}}{2m}
-\frac{\vec{p}\,'^{\,2}}{2m} + ...
\label{Dc}
\end{equation}
On the other hand, corrections which arise from the expansion of Eq. (\ref{Bb}) 
up to terms of order $\frac{1}{\omega^3 }$, are consistent with Galilean 
invariance in the standard non-relativistic approximation where 
$e_p$ is replaced by $ m+ \frac{p^2}{2m}$. This supposes that the contributions 
of both two terms in Eq. (\ref{Ba}) are taken into account. It can indeed be 
shown that the complete numerator of the second term in the expansion given by 
Eq. (\ref{Ca}) is then independent of the total momentum, as it follows from the 
equality
$$E_{c.m.}+\frac{(\vec{p_1}+\vec{p_2})^{\,2}}{4m} -\frac{\vec{p_1}^{\,2}}{4m} 
-\frac{\vec{p_2}^{\,2}}{4m} -\frac{\vec{p_1}\,'^{\,2}}{4m} 
-\frac{\vec{p_2}\,'^{\,2}}{4m} =
E_{c.m.}-\frac{\vec{p}^2}{2m}-\frac{\vec{p}\,'^{\,2}}{2m},$$
where $E_{c.m.}$ is the energy in the center of mass (simply written $E$ 
elsewhere in the text). The corrections under discussion are therefore 
completely consistent with a non-relativistic approach. 

Corrections of higher order in $\frac{1}{\omega }$ depend on the total momentum 
of the system. Contrary to the previous ones, they have a relativistic 
character, accounting in particular for the fact that two events that occur 
in a given order in one frame may occur in a different order in another one. 
This effect is also closely related to the occurence of two different terms 
in Eq. (\ref{Ba}), which reduce to a single one in the center of mass system.

Being justified in the Born approximation, the one-boson exchange instantaneous 
approximation cannot correctly account for higher order (off-shell) effects. 
How large these ones are  will be discussed in the next sections where binding
energies are compared with corresponding ones of more complete 
approaches. The above statement also applies to the many quasi-potential 
approaches that have been considered in the literature \cite{GROS,BSLT}, 
often with the idea to provide some link to the Bethe-Salpeter equation. 
Frequently, they differ by the two-particle propagator while using an 
interaction similar to the so-called instantaneous approximation. 
Little attention has been given to a significant modification of the 
interaction, like the one provided by a renormalization as given by Eq. 
(\ref{Ch}). This one, as will be seen below, is essential to reveal the close 
relationship with the Bethe-Salpeter equation. Notice that, in the domain of 
hadronic physics, this renormalization is accounted for correctly in different 
ways at the two-pion level in the Paris  \cite{PARI} and full Bonn \cite{BONN} 
models of the NN interaction, but treated phenomenologically in most other ones. 

Energy-dependent corrections beyond the instantaneous approximation, which  we 
are interested in here, have different names, depending on the domain where they 
are employed: norm corrections in the case of mesonic exchange currents 
\cite{CHEM}, dispersive effects in scattering processes involving composite 
systems \cite{MAHA} or retardation effects in two-body scattering \cite{BONN}. 
It is remarked that retardation effects sometimes represent different kind of 
corrections, involving terms of the order 
$\left(\frac{\Delta E}{\omega}\right)^2$
\cite{FRIA}, rather than the ones retained here, that involve terms of the 
order  $\frac{\Delta E}{\omega}$.
As already mentioned, the former ones are of the order 
$\left(\frac{1}{m}\right)^2$, and have a typical relativistic character. 

\subsection{Some notes on abnormal states}

It is a common feature of all non-relativistic reduction formalisms of the
Bethe-Salpeter equation, that the original relative time degree of freedom is
somehow eliminated. One is thereby faced with the question whether an
essential dynamical ingredient is lost in this step and if yes, how to recover 
it. The main motivation of this work came from the conviction that the dynamics 
involving the relative time can indeed be accounted for in a non-relativistic 
scheme. If this is the case, also excited states should be well described in an
effective approach. This is why we shall put some emphasis on the simultaneous
description of ground and excited states when we present our results in the next
section. 

If we claim that the dynamics of the time variable is well described in a
non-relativistic approach, one may immediately ask about states,
that are intrinsically connected to this fourth variable, that are the so-called
abnormal states. Being excitations in the fourth dimension, it is difficult to
imagine describing them in a three-dimensional formalism. It was noted however
\cite{BIJT} that just this may be achieved by an energy-dependent potential.
The existence of ``extra'' solutions (i.e. solutions that do not have an
analogue when the energy dependence in the potential is neglected)
to the Schr\"odinger equation was analytically demonstrated in 
ref. \cite{KHEL}. The authors considered the equivalent of our Eq. (\ref{Be}) 
for the zero-mass boson case ($\mu=0$). The key point was the observation, that 
the asymptotic behaviour of the potential in configuration space changes 
discontinuously at large distances if the energy dependence is kept. 
The potential then takes the asymptotic form
\begin{equation}
\label{asypot}
V_E(r)|_{Er\gg1} \sim -\frac{2\alpha}{Er^2},
\end{equation}
instead of the usual Coulomb potential $V_{E=0}(r)=-\frac{\alpha}{r}$. Thereby,
a new class of solutions appears, characterized by a radial wave function 
with an infinity of zeroes. It was also noted, that these solutions exist only
for large values of the coupling constant, which is just what happens in the
relativistic analogue of this case, the Wick-Cutkosky model \cite{WICK}.

It seems possible therefore to account for abnormal states even in a 
three-dimensional formalism. Unfortunately, neither of the effective
interaction potentials that we shall use, Eq. (\ref{Cy}) and Eq. (\ref{Ci}), 
even though
they are supposed to account for a part of the energy dependence of the original
potential that they were derived from, will be able to produce these
supplementary states. The reason is, that they both have the same asymptotic
behaviour for large distances in configuration space as the original Coulomb
potential, so the argument of ref. \cite{KHEL} does not apply. This shows the
nonperturbative nature of these states and the necessity to keep track of the 
full energy dependence of the original potential if one wants to account for
them. However, from a calculational point of view, it seems to be difficult 
to identify these solutions numerically, since there is no obvious criterion to
distinguish them from normal ones. For that reason and in order to render a 
comparison meaningful for all approaches considered here, we shall only
investigate normal states in this work.

\section{Results in ladder approximation}

\begin{figure}[ht]
\begin{center}

\mbox{\rule[0cm]{0mm}{1cm}
\psfig{file=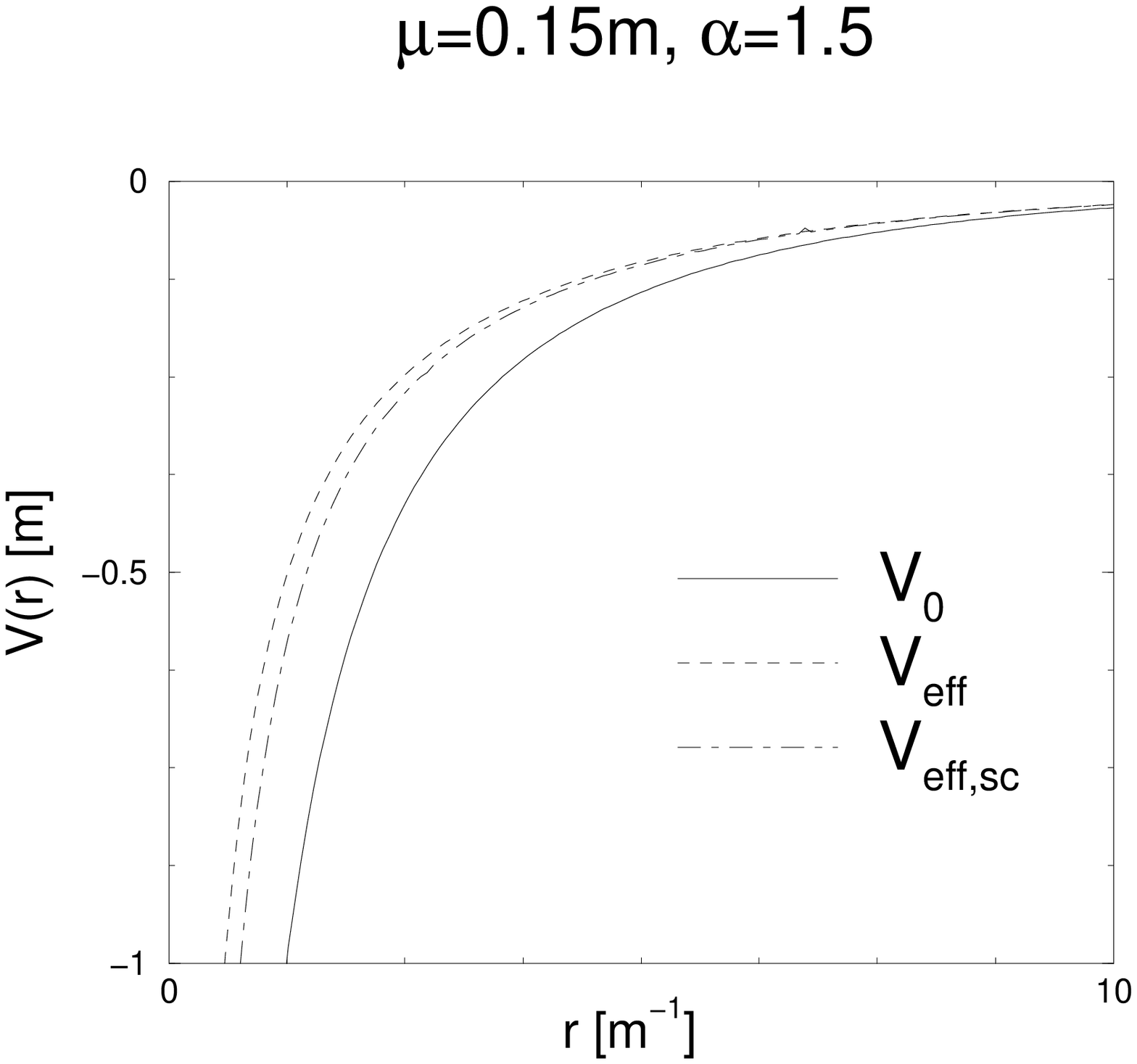,width=14em,height=12em}\mbox{\hspace{2em}}
\psfig{file=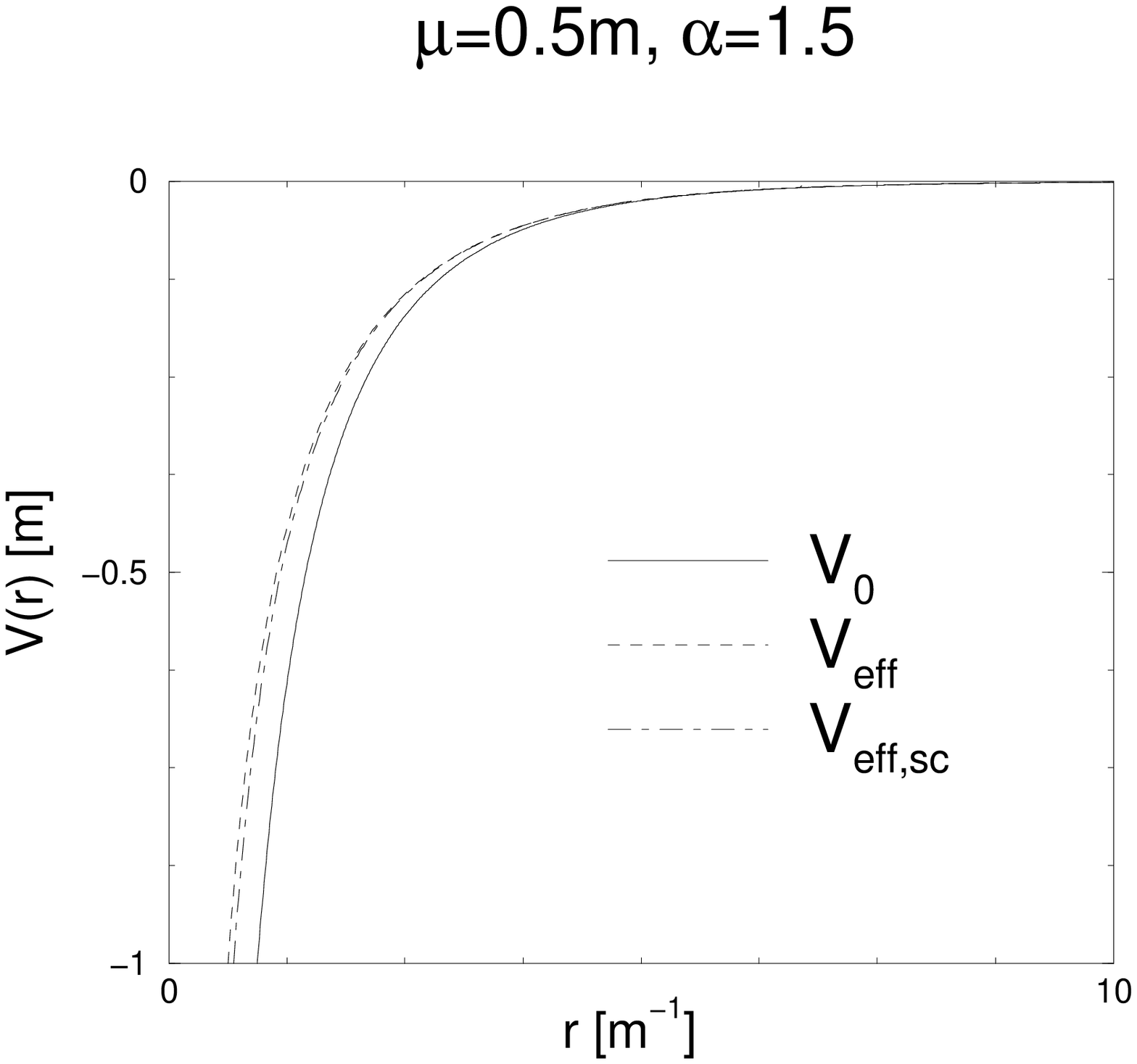,width=14em,height=12em} }
\vspace*{1cm}

\mbox{\rule[0cm]{0mm}{1cm}
\psfig{file=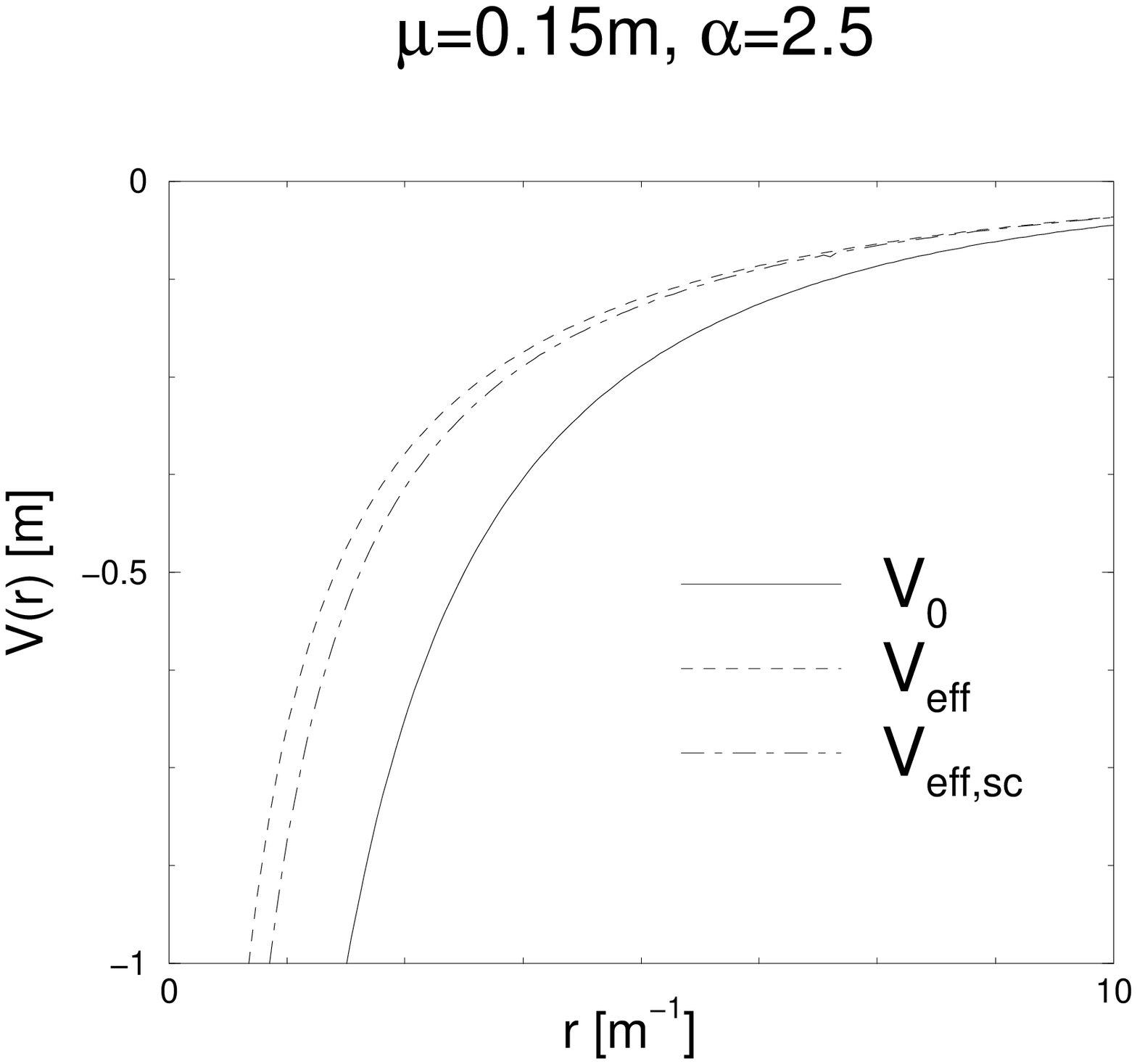,width=14em,height=12em}\mbox{\hspace{2em}}
\psfig{file=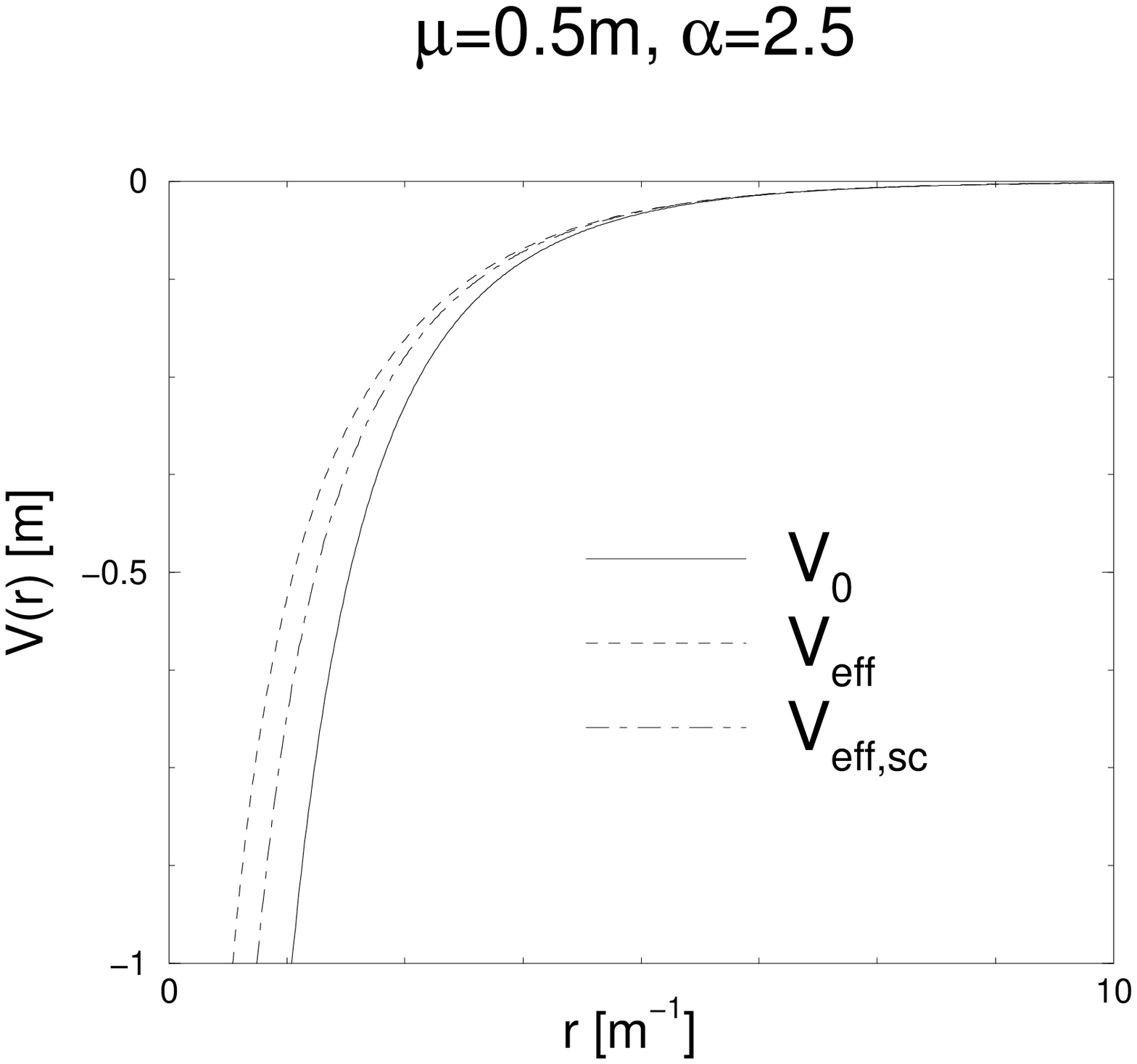,width=14em,height=12em}}
\caption{\label{fig4}Representation of non-relativistic potentials for two 
values of $\alpha$ and two values of $\mu$. $V_0$ is the instantaneous 
potential of Eq. (\protect\ref{Cd}), $V_{eff}$ is the effective potential of 
Eq. (\protect\ref{Ci}) and $V_{eff,sc}$ gives the self-consistent, 
effective potential of Eq. (\protect\ref{Cy}).}
\end{center}
\end{figure}

We present here binding energies calculated in ladder approximation using the
different approaches described in the last section. Explicitly, these ones 
include:
\begin{itemize}
\item the Bethe-Salpeter (B.S.) equation, Eq. (\ref{Aa}), 
together with an interaction 
kernel given by Eq. (\ref{Ab}) (ladder approximation),
\item a semi-relativistic equation, Eq. (\ref{Bc}), with  an energy-dependent 
interaction ($V_E$) given by Eq. (\ref{Bb}), 
\item the standard non-relativistic Schr\"odinger equation, Eq. (\ref{Cg}), 
together with an effective interaction ($V_{eff,sc}$) given by 
Eq. (\ref{Cy}), which accounts for energy-dependent effects 
in a self-consistent calculation,
\item the same equation, but with an effective interaction ($V_{eff}$)
as given by Eq. (\ref{Ci}), accounting for  
energy-dependent effects only at the lowest order, 
\item the same equation in the so-called instantaneous approximation (IA) with
an interaction given by Eq. (\ref{Cd}).
\end{itemize}
In Fig. \ref{fig4}, we plot the potential functions of the last three of these 
approaches in configuration space with the box contribution included for two 
values of the coupling constant and two values of the boson mass. It can be seen 
that the effective potentials are significantly smaller than the 
instantaneous approximation, especially at small distances, the effect being 
larger for $V_{eff}$  than for its self-consistent version, $V_{eff,sc}$. 

In the following, we give results for the above two different boson masses as 
well as different angular momenta. The aim is to show that binding energies
are relatively close to each other for the four first approaches and 
significantly depart from those obtained in the instantaneous one-boson exchange 
approximation. In the second, third and fourth approaches, the results with and 
without the contribution of the box diagram (Fig. \ref{fig3} a) are considered, 
so that a comparison with the Bethe-Salpeter approach is fully relevant at order 
$\left(\frac{1}{m}\right)^0$.

\subsection{Finite-mass boson case}

\begin{sidewaystable}
\caption{Binding energies calculated in the ladder approximation for massive
scalar bosons (in units of the constituent mass $m$). Results are presented here 
as a function of the coupling constant $\alpha$ for the ground state ($l=0$) 
and two boson masses ($\mu=0.15\,m$ and $\mu=0.5\,m$). Five cases have been 
considered: the Bethe-Salpeter equation (B.S., Eqs. (\ref{Aa}, \ref{Ab})), an 
energy-dependent model ($V_E$, Eqs. (\ref{Bb}, \ref{Bc})), an effective 
non-relativistic interaction ($V^{nr}_{eff,sc}$, Eqs. (\ref{Cg}, \ref{Cy})), a 
simplified version of this effective interaction  ($V^{nr}_{eff}$, Eqs. 
(\ref{Cg}, \ref{Ci})) and, slightly apart, the so-called instantaneous 
approximation (IA$^{nr}$, Eqs. (\ref{Cg}, \ref{Cd})). 
In the second,  third and fourth cases, the binding 
energies calculated without the contribution of the box diagram are 
given between parentheses. The superscript {\it nr} reminds that 
the calculations are performed with an interaction derived at the lowest order, 
$\left(\frac{1}{m}\right)^0$.}
\label{tab1}
\renewcommand{\arraystretch}{1.1} 
\begin{tabular*}{\textheight}{@{\extracolsep{\fill}}cccccccc}
\hline  
\rule{0pt}{3.3ex}
  $\mu/m$ & $\alpha$   & ${\rm B.S.}$ & $V_E \;(-{\rm box}) $& 
$V^{nr}_{eff,sc} \;(-{\rm box}) $  & $V^{nr}_{eff} \;(-{\rm box}) $ 
  & \hspace*{0.5cm}& ${\rm IA}^{nr} $ \\ [1.ex]
\hline
$$    & $0.50$ & 0.0059& 0.0052 (0.0045)&0.0062 (0.0051) & 0.0055 (0.0040) & & 
0.013  \\ [0.ex]
$$    & $0.75$ & 0.023 &  0.020 (0.017) &0.023 (0.020) & 0.020 (0.014)   & & 
0.056  \\ [0.ex] 
$.15$ & $1.00$ & 0.047 &  0.039 (0.034) &0.048 (0.039) & 0.038 (0.026)   & & 
0.129  \\ [0.ex]
$$    & $1.25$ & 0.076 &  0.061 (0.052) &0.076 (0.063) & 0.059 (0.037)   & & 
0.233  \\ [0.ex]
$$    & $1.50$ & 0.109 &  0.084 (0.073) &0.108 (0.088) & 0.081 (0.047)   & & 
0.367  \\ [1.ex]    
$$    & $1.50$ & 0.0125&  0.0071 (0.0051)&0.0144 (0.0090)& 0.0098 (0.0028) & & 
0.093  \\ [0.ex] 
$$    & $1.75$ & 0.028 &  0.017 (0.013) &0.031 (0.021) & 0.023 (0.008)   & & 
0.180  \\ [0.ex] 
$.50$ & $2.00$ & 0.049 &  0.030 (0.024) &0.053 (0.037) & 0.038 (0.014)   & & 
0.296  \\ [0.ex]
$$    & $2.25$ & 0.073 &  0.045 (0.036) &0.078 (0.056) & 0.056 (0.022)   & & 
0.442  \\ [0.ex]
$$    & $2.50$ & 0.100 &  0.061 (0.050) &0.106 (0.077) & 0.077 (0.030)   & & 
0.619  \\ [1.ex]
 
\hline
\end{tabular*}
\end{sidewaystable}

\begin{sidewaystable}
\caption{Same as Table \protect\ref{tab1} for excited states with $l=0^{*}$ 
and $l=1$. Bars 
indicate that there is either no bound state or that its binding energy, being 
too small, cannot be determined reliably.}
\label{tab2}
\renewcommand{\arraystretch}{1.1} 
\begin{tabular*}{\textheight}{@{\extracolsep{\fill}}ccccccccc}
\hline
\rule{0pt}{3.3ex}
  $\mu/m$&$l$&$\alpha$   & ${\rm B.S.}$ & $V_E \;(-{\rm box}) $ & 
$V^{nr}_{eff,sc} \;(-{\rm box}) $   &$V^{nr}_{eff} \;(-{\rm box}) $   & 
\hspace*{0.5cm}& ${\rm IA}^{nr} $ \\ [1.ex]
\hline
 $     $&$   $&$ 2.00 $ & 0.005 & 0.003 (0.002) &0.006 (0.004) &0.003 
\makebox[3em]{( ----- )}       
    & & 0.054 \\ [0.ex]
 $     $&$   $&$ 2.25 $ & 0.010 & 0.006 (0.004) &0.011 (0.007) &0.005 
\makebox[3em]{( ----- )}          
& & 0.086     \\ [0.ex]
 $   $&$0^{*}$&$ 2.50 $ & 0.015 & 0.009 (0.007) &0.016 (0.011) &0.009 (0.001)   
& & 0.124  \\ [0.ex]
 $     $&$   $&$ 2.75 $ & 0.022 & 0.013 (0.010) &0.023 (0.015) &0.013 (0.002)   
& & 0.170  \\ [0.ex]
 $     $&$   $&$ 3.00 $ & 0.029 & 0.018 (0.014) &0.030 (0.021) &0.018 (0.003)     
 & & 0.223      \\ [0.ex]
 $0.15 $                &        &               & &             & &        \\ 
 $     $&$   $&$ 2.00 $ & 0.003 & 0.002 (0.001) &0.003 (0.001) &0.002 
\makebox[3em]{( ----- )}         
  & & 0.042  \\ [0.ex]
 $     $&$   $&$ 2.25 $ & 0.009 & 0.006 (0.003) &0.009 (0.005) &0.006 (0.001)   
& & 0.073  \\ [0.ex]
 $     $&$ 1 $&$ 2.50 $ & 0.015 & 0.011 (0.007) &0.015 (0.009)&0.011 (0.003)   
 & & 0.111  \\ [0.ex] 
 $     $&$   $&$ 2.75 $ & 0.022 & 0.016 (0.011) &0.022 (0.014)&0.016 (0.005)   
 & & 0.156  \\ [0.ex]
 $     $&$   $&$ 3.00 $ & 0.030 & 0.022 (0.016) &0.029 (0.020)&0.022 (0.007)   
 & & 0.209  \\ 
[1.ex]
 \hline
 $     $&$   $&$ 9    $ & 0.022  & \makebox[2.3em]{( --- )} \makebox[3em]{( 
----- )} &0.031 (0.012) 
&0.017 
\makebox[3em]{( ----- )}  & & 1.787      \\ [0.ex] 
 $     $&$   $&$ 10   $ & 0.042  & 0.004 (0.001) & 0.053 (0.024) & 0.035 
\makebox[3em]{( ----- )}  & & 2.494     \\ [0.ex]
        &\raisebox{1.5ex}[0pt]{$0^{*}$}
              &$ 11   $ & 0.067  & 0.012 (0.005) & 0.079 (0.040) & 0.057 
\makebox[3em]{( ----- )}  & & 3.324      \\ [0.ex]
 $     $&$   $&$ 12   $ & 0.096  & 0.022 (0.011) & 0.108 (0.058) & 0.084 
\makebox[3em]{( ----- )}  & & 4.279      \\ [0.ex]
 $ 0.5 $                &        &               &  &            & &        \\
 $ $    &$   $&$ 9    $ & 0.012 & \makebox[2.3em]{( --- )} \makebox[3em]{( ----- 
)} & 
0.014  \makebox[3em]{( ----- )} & \makebox[2.3em]{( --- )} \makebox[3em]{( ----- 
)} & & 1.627  \\ [0.ex] 
 $ $   &$   $&$ 10   $ & 0.037  & 0.003 \makebox[3em]{( ----- )} &0.039 (0.008) 
&0.022 \makebox[3em]{( ----- )}  & & 2.327  \\ [0.ex]
        &\raisebox{1.5ex}[0pt]{$1$}
              &$ 11   $ & 0.067  & 0.015 (0.002) &0.067 (0.026) &0.046 
\makebox[3em]{( ----- )}  & & 3.150     \\ [0.ex]
 $ $    &$   $&$ 12   $ & 0.101  & 0.031 (0.012) &0.099 (0.046) &0.075 
\makebox[3em]{( ----- )}  & & 4.099      \\ [1.ex]
 
\hline
\end{tabular*}
\end{sidewaystable}

Calculations have been performed with boson masses $\mu=0.15\,m$ and 
$\mu=0.5\,m$, a choice that was made in earlier works \cite{NIEU,FRED}. In the 
instantaneous approximation, these results are not independent. The ratio of the 
binding energy and the squared boson mass is a unique function of the ratio of 
the coupling constant and the boson mass: 
\begin{equation}
\frac{m\,E_{be}}{ \mu^2}= f \left(\frac{\alpha m }{ \mu}\right).
\label{Ea}
\end{equation}
For the coupling constant, $\alpha=\frac{g^2}{4\pi}$, we considered ranges 
appropriate for the states under consideration, in order to provide a 
significant insight on the variation of the binding energy with the strength of 
the interaction. Some of the binding energies may be quite large, questioning 
their physical relevance. However, this should not spoil the comparison of the 
different approaches we are considering, which is our main intent. For the 
angular momentum, we obviously considered the ground state with $l=0$ 
(Table \ref{tab1}), but also the first excited states with $l=0$ (denoted 
$0^{*}$) and $l=1$. The energies for the latter ones, which are degenerate in 
the case of zero-mass boson exchange, are presented in Table \ref{tab2}. Their 
examination may provide useful information on the role of the combined effect of 
the coupling strength and the boson mass.

Examination of Tables \ref{tab1} and \ref{tab2} shows that results from the 
first four approaches are relatively close to each other and definitively depart 
from the instantaneous-approximation results. While this qualitative conclusion 
is not essentially affected by the contribution of the box diagram, this one is 
quite important in showing the approximate quantitative equivalence of the four 
first approaches in all cases. 

When comparing results of $V^{nr}_{eff,sc}(r)$ and $V^{nr}_{eff}(r)$,
it is noted that the self-consistency condition of Eq. (\ref{Cy}) 
plays a non-negligible role. It makes the results more consistent with what we 
expect from possible corrections (see the discussion at the end of the section).

\begin{figure}[htb]
\begin{center}
\mbox{\psfig{file=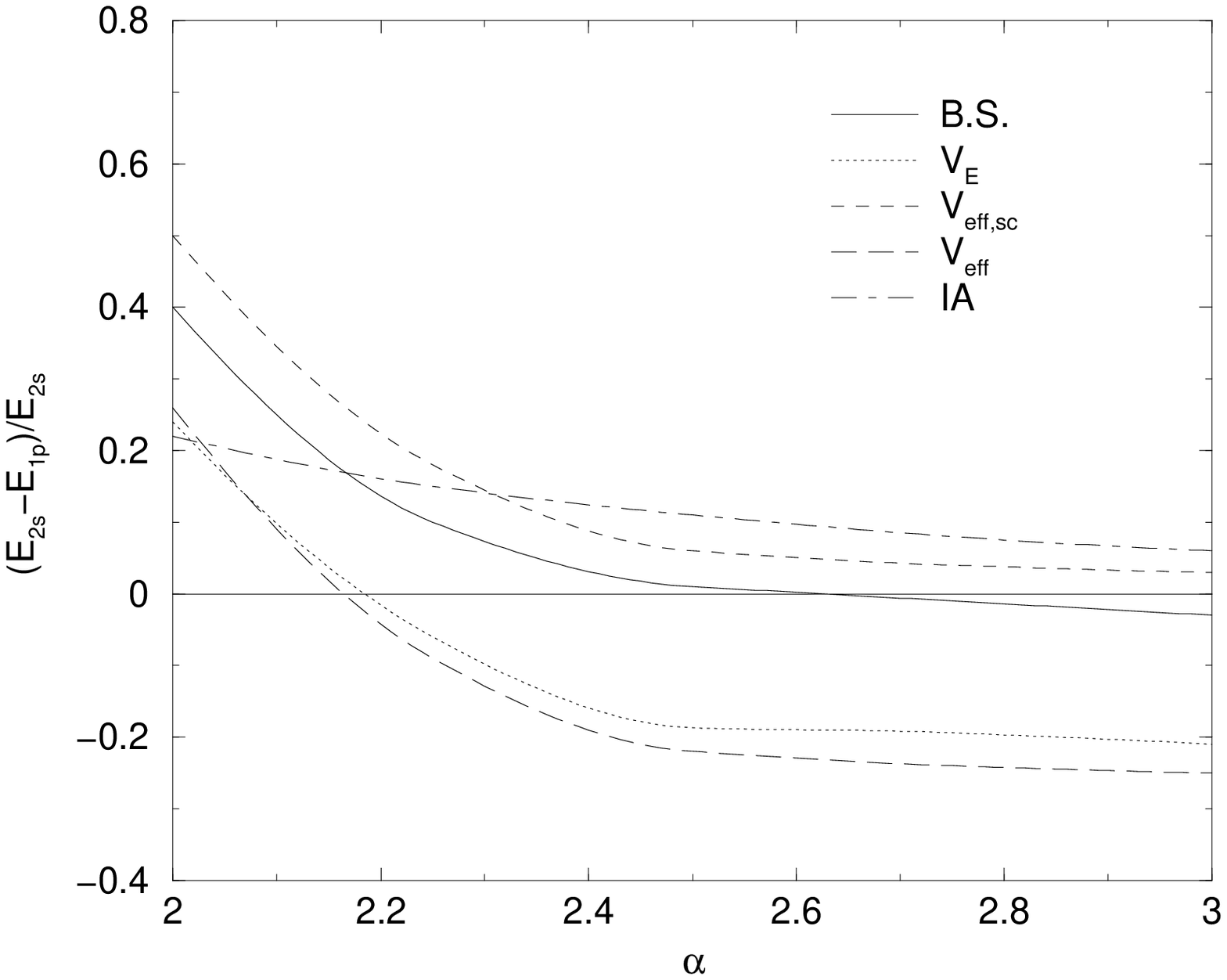,width=15em}\mbox{\hspace{2em}}
\psfig{file=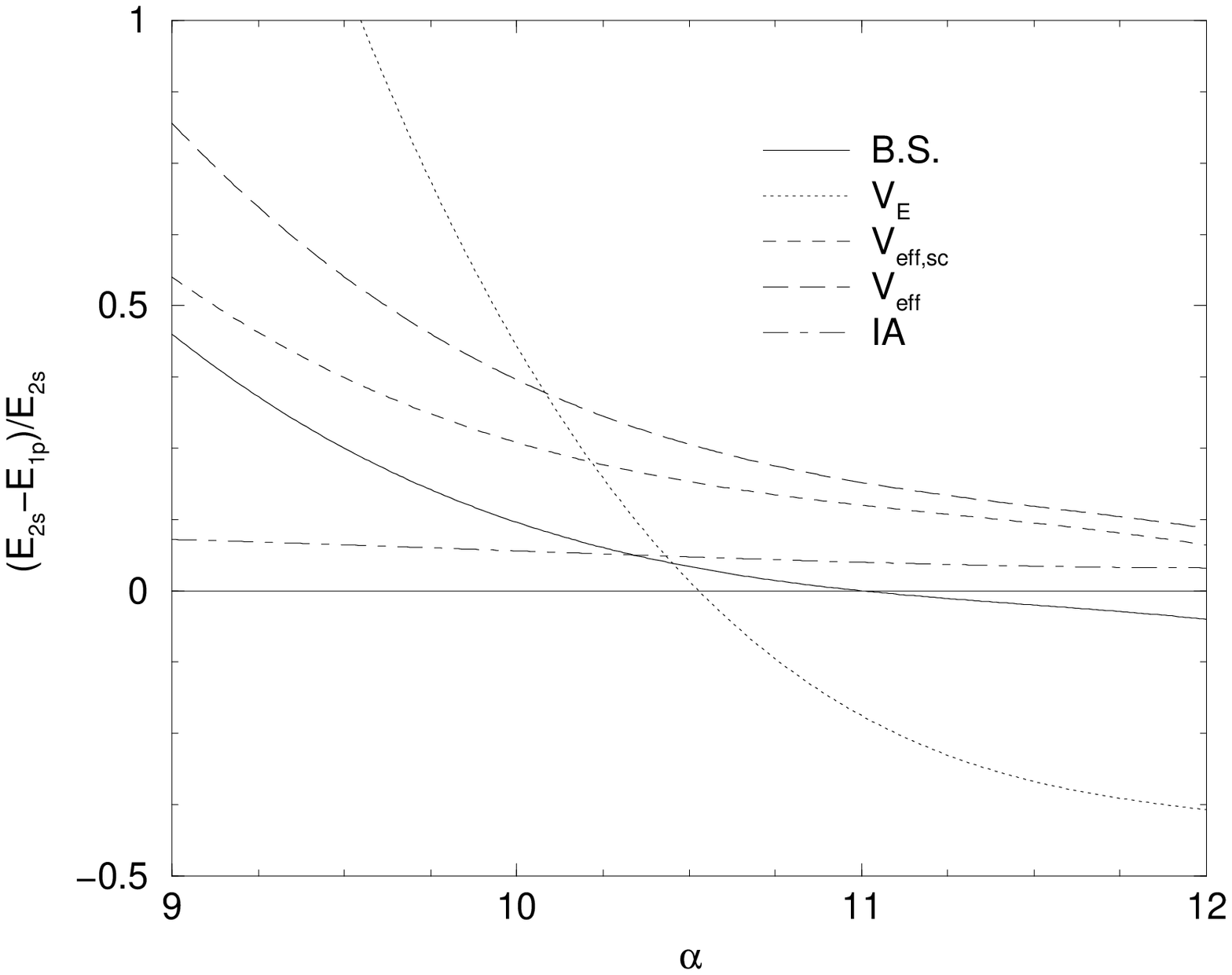,width=15em}}  
\end{center}
\caption{\label{fig5}
Relative ordering of binding energies of the first radial (2s) 
and orbital (1p) excited states for different approaches. 
The figure on the left corresponds
to the case $\mu=0.15\, m$, the one on the right to $\mu=0.5 \,m$.}
\end{figure}

Concerning the binding energies of the lowest $l=1$ state and the first excited 
state with $l=0$, some track of the degeneracy expected for a Coulomb potential 
can be seen in Table \ref{tab2} (for states that are not too weakly bound). 
It is interesting however to observe the relative order of these 
states in the various approaches. In the instantaneous approximation the total 
energy of the first excited state with $l=0$ is always smaller than the one of 
the $l=1$ state for all values of the coupling constant, as expected from a 
theorem involving the concavity of the potential \cite{BMG}. The situation is 
however different in the energy-dependent model, and in the Bethe-Salpeter case, 
where a crossing of the 2s and 1p states occurs. This peculiarity is illustrated 
in Fig. \ref{fig5}, where we plot the ratio of binding energies of the two 
states for the various approaches. This feature, together with the principal 
possibility of the description of abnormal states, as discussed earlier, shows 
the dynamical importance of the energy dependence of the interaction for 
describing also excited states. 

\subsection{Zero-mass boson case}
Binding energies for the zero-mass boson case have been considered in various 
works from different viewpoints (see for instance refs. \cite{SILV,BILA,JI}). 
Those for the Wick-Cutkosky model obtained with the Bethe-Salpeter equation in 
ref. \cite{SILV} are reminded here,  complemented by a few more values of the 
coupling constant, $\alpha$. Binding energies calculated with the 
energy-dependent interaction, $V_E$, may be compared to those obtained using the 
memory-function approach in ref. \cite{BILA}. The latter one includes higher 
order corrections in $\frac{1}{m}$, but neglects the contribution of the box 
diagram (Fig. \ref{fig3} a). The two last sets of numbers considered here 
correspond to binding energies calculated in a non-relativistic approach with an 
effective Coulomb interaction accounting for energy-dependent effects, and the 
bare Coulomb interaction as a reference. The binding energy for these cases 
is then given by the  standard analytic expression:
\begin{equation}
|E_{be}|=\frac{\alpha^2}{4n^2}\;\;\;
\left({\rm or} \; \frac{\alpha^2_{eff,sc}}{4n^2}\right),
\label{Fa}
\end{equation}
where the relation between $\alpha_{eff,sc}$ and $\alpha$, first studied in ref. 
\cite{AMGH}, is given in the simplest case by:
\begin{equation}
\alpha_{eff,sc} =  \frac{\alpha}{ 1 +  \alpha \, J}
 =  \alpha (1 - \alpha_{eff,sc} \, J),\;\;{\rm with}\;\; 
 J=\frac{2}{\pi} \, \int_0^{\infty} \frac{dy \, j_0(y)}{\alpha_{eff,sc}+y}.
 \label{Fb}
\end{equation}
This equation can also be obtained from Eq. (\ref{Cy}) for the zero-mass boson
case. It can be improved by incorporating the contribution of the box diagram, 
in which case it gets modified as follows: 
\begin{equation}
\alpha_{eff,sc} =  \frac{\alpha(1+\alpha J^b)}{ 1 +  \alpha \, J}
 =  \alpha (1+\alpha J^b - \alpha_{eff,sc} \, J),
 \label{Fc}
\end{equation}
where $J$ is still given by the integral in Eq. (\ref{Fb}). The 
quantity $J^b$, which involves the contribution of the box diagram 
(Fig. \ref{fig3} a), is given by:
\begin{eqnarray}
 J^b=\frac{1}{\pi^2}\,\int_0^{\infty}dx \, j_0(x)\int_0^{\infty}\frac{dy 
\;\;\;x}{y(\alpha_{eff,sc}+y)} \,\nonumber \hspace{5cm} \\ \times
 \log\left(\frac{(\alpha_{eff,sc}+|x+y|)(\alpha_{eff,sc}+y+|x-y|)}{
 (\alpha_{eff,sc}+|x-y|)(\alpha_{eff,sc}+y+|x+y|)} \right),
 \label{Fd}
\end{eqnarray}
where the second integral becomes equal to $4 \, \log 2$ in the limit of  
small $\alpha_{eff,sc}$. It is calculated from an expression similar to Eq. 
(\ref{Bf}) where the terms $E-\frac{p^2}{m}$ appearing in the full expression 
have been kept and approximated by $-\frac{\alpha_{eff,sc}}{r}$. This is 
consistent with the calculation of $J$, allowing one to factorize a local 
Coulomb-type contribution. While the relation of $J^b$ to the two-boson 
exchange can be traced back without too much difficulty, this is not so 
obvious for the quantity $J$ which was obtained by accounting for off-shell 
effects arising from one-boson exchange. However, it can be made more 
transparent by rewriting $J$ in a more complicated form that is  formally 
similar to Eq. (\ref{Fd}). Using the identity:
$$\sin(y)=\frac{1}{\pi} \, \int_0^{\infty}dx \, \sin(x)
\, \, \log \left( \frac{|x+y|}{|x-y|} \right), $$
one gets: 
\begin{equation}
 J=\frac{2}{\pi^2} \, \int_0^{\infty}dx \, j_0(x)\int_0^{\infty}\frac{dy 
\;\;\;x}{y(\alpha_{eff,sc}+y)} \, \log \left( \frac{|x+y|}{|x-y|} \right ).
\end{equation}

\begin{figure}[tb]
\begin{center}
\mbox{\psfig{file=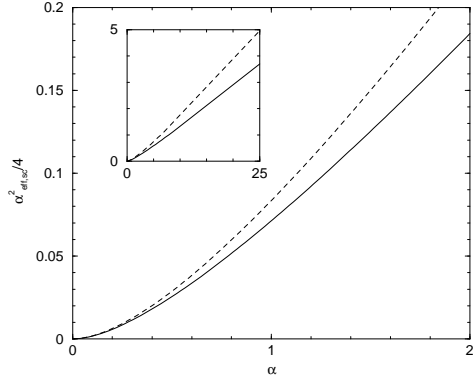,width=15em}}  
\end{center}
\caption{\label{fig6}
Representation of $\frac{\alpha_{eff,sc}^2}{4}$  as a function of the bare 
coupling $\alpha$. The solid line corresponds to the solution of Eq.
(\protect\ref{Fb}), the dashed line to the one of Eq. (\protect\ref{Fc}).
The inset shows the linear asymptotic behaviour for large $\alpha$.}
\end{figure}

Curves showing $\alpha_{eff,sc}$ as a function of the coupling 
$\alpha$ are given in Fig. \ref{fig6} for the different cases corresponding to 
Eqs. (\ref{Fb}) and (\ref{Fc}). More precisely, we represent the quantity 
$\frac{\alpha_{eff,sc}^2}{4}$ as a function of $\alpha$. This one offers the 
advantage of being directly comparable to the binding energy obtained for the 
lowest $l=0$ state in a non-relativistic Coulomb-like potential, 
$-\frac{\alpha_{eff,sc}}{r}$ (we set $m=1$ here). It can be as well 
compared to the product of the binding energy by the factor $n^2$ where $n$ is 
the quantum number relative to the excited states obtained in such a potential. 
It can also be compared to the quantity $\frac{4m^2-E^2}{4m}$ (possibly 
multiplied by the factor $n^2$) which replaces the above binding energy in some 
relativistic approaches while the equations to be solved keep the same form. For 
large couplings, $\alpha_{eff,sc}$ varies like $\sqrt{\alpha}$, which can 
be checked from Eq. (\ref{Fb}) or equivalently from Eq. (\ref{Cy}) in the limit 
of a large effective potential.  One gets the asymptotic relation $ 
\alpha_{eff,sc}^2 \rightarrow  \frac{2}{\pi} \, \alpha$ for one-boson exchange, 
Eq. (\ref{Fb}) and $\alpha_{eff,sc}^2 \rightarrow  \frac{2}{\pi(\sqrt{3}-1)} \, 
\alpha $ when also including two-boson exchange, Eq. (\ref{Fc}). The 
corresponding result for the Wick-Cutkosky model would be  $
 \alpha_{eff,sc}^2 \rightarrow \frac{4}{\pi} \, \alpha $.

\begin{sidewaystable}
\caption{\label{tab3}
Binding energies calculated in the ladder approximation for zero-mass 
scalar bosons (Wick-Cutkosky model).  Results (in units of the constituent mass 
m) are presented here as a function of the coupling constant $\alpha$ for the 
ground state ($l=0$) as well as excited states ($l= 0^{*}, 1, 0^{**}, 1^{*}, 
2$). Four cases from the previous tables have been considered here: the 
Bethe-Salpeter equation (B.S.), an energy-dependent model ($V_E$), an effective 
interaction $V^{nr}_{eff,sc}$ and, slightly apart, the so-called 
instantaneous approximation(IA$^{nr}$). For $V_E$ the binding energies for 
excited states, which are not degenerate in this case, given in the table 
correspond to the highest orbital momentum $l$ only. Bars indicate that the 
corresponding binding energy is too small to be determined reliably using the 
same numerical method as for the other numbers in the same column.}
\renewcommand{\arraystretch}{1.1} 
\begin{tabular*}{\textheight}{@{\extracolsep{\fill}}clccccc}
\hline
\rule{0pt}{3.3ex}
  $\alpha$ & $l$   & ${\rm B.S.}$ & $V_E \;(-{\rm 
box}) $ & $V^{nr}_{eff,sc}\; (-{\rm box}) $   & \hspace*{0.5cm}& ${\rm IA}^{nr} 
$ 
\\ 
[1.ex]
\hline
$$          & $0,$              & 1.274-5 & \makebox[3em]{( ----- )} & 1.274-5 
(1.266-5) & & 1.331-5\\ 
[0.ex] 
$1/137.036$ & $0^{*},1,$        & 0.318-5 & \makebox[3em]{( ----- )} & 0.318-5 
(0.316-5) & & 0.333-5\\ 
[0.ex]
$$          & $0^{**},1^{*},2,$ & 0.141-5 & \makebox[3em]{( ----- )} & 0.141-5 
(0.140-5) & & 0.148-5\\ 
[1.ex] 
$$	    & $0,$	        & 1.849-3 & \makebox[3em]{( ----- )} & 1.859-3 
(1.770-3) & & 2.500-3\\ 
[0.ex] 
$0.10$      & $0^{*},1,$        & 0.465-3 & \makebox[3em]{( ----- )} & 0.465-3 
(0.442-3) & & 0.625-3\\ 
[0.ex]
$$          & $0^{**},1^{*},2,$ & 0.207-3 & \makebox[3em]{( ----- )} & 0.207-3 
(0.197-3) & & 0.278-3\\ 
[1.ex] 
$$          & $0,$              & 2.867-2 & \makebox[3em]{( ----- )} & 2.883-2 
(2.547-2) & & 6.250-2\\ 
[0.ex]
$0.50$      & $0^{*},1,$        & 0.727-2 & \makebox[3em]{( ----- )} & 0.721-2 
(0.637-2) & & 1.562-2\\ 
[0.ex]
$$          & $0^{**},1^{*},2,$ & 0.324-2 & \makebox[3em]{( ----- )} & 0.320-2 
(0.283-2) & & 0.694-2\\ 
[1.ex]
$$          & $0,$              & 0.842-1 & 0.725-1 (0.637-1) & 0.832-1 
(0.711-1) & & 2.50-1 \\ 
[0.ex]
$1.00$      & $0^{*},1,$        & 0.213-1 & 0.202-1 (0.174-1) & 0.208-1 
(0.178-1) & & 0.62-1 \\ 
[0.ex]
$$          & $0^{**},1^{*},2,$ & 0.095-1 & 0.093-1 (0.081-1) & 0.092-1 
(0.079-1) & & 0.28-1 \\ 
[1.ex]
$$          & $0,$              & 1.545-1 & 1.239-1 (1.071-1) & 1.494-1 
(1.247-1) & & 5.62-1 \\ 
[0.ex]
$1.50$      & $0^{*},1,$        & 0.388-1 & 0.353-1 (0.300-1) & 0.373-1 
(0.312-1) & & 1.41-1 \\ 
[0.ex]
$$          & $0^{**},1^{*},2,$ & 0.173-1 & 0.164-1 (0.138-1) & 0.166-1 
(0.139-1) & & 0.62-1 \\ 
[1.ex]
 
\hline
\end{tabular*}
\end{sidewaystable}

The different results for the binding energies are given in Table \ref{tab3}. 
The results of $V^{nr}_{eff}$ relative to those of $V^{nr}_{eff,sc}$ are very 
similar to the corresponding ones of Table \ref{tab1}, so we do not give the 
numbers here. In the case of $V_E$, the small binding energies obtained here for 
small couplings cannot be reliably determined by the variational procedure that 
we employ. Only values for relatively large couplings are given here, in order 
to allow for a comparison of different approaches. Note also, that for the 
energy-dependent interaction,  $V_E$, binding energies for excited states are 
not degenerate, as it is the case for  all the other models considered here, 
even though the splitting is expected to be small. The numbers that we present 
in Table \ref{tab3} for this case correspond to the state with the highest value 
of angular momentum, $l$, in that row.

As for the massive boson case, the three first calculations are close to each 
other and depart from the instantaneous approximation results. Again, the 
contribution of the box diagram, without affecting the above conclusion, 
si\-gni\-ficantly helps in narrowing the gap between the Bethe-Salpeter approach 
(B.S.) and those accounting for energy-dependent effects, $V_E$  and 
$V_{eff,sc}$. The results for the Bethe-Salpeter case and the effective 
interaction are very close to each other, especially if one compares them to 
those obtained with the energy-dependent interaction, $V_E$. This is due to the 
cancellation of different relativistic corrections that will be discussed in the 
next subsection. It is also noticed that the discrepancy between the B.S. and 
$V_E$ results decreases with $l$. As this difference mainly originates from 
Z-type diagrams, which produce a $\frac{1}{r^2}$ contribution to the interaction 
in the non-relativistic limit, it is expected that its effect be reduced by the 
centrifugal barrier, providing an explanation for the above observation.

Due to the way the effective potential is derived, it evidences the degeneracy 
pattern of the Wick-Cutkosky model in the ladder approximation, as does the IA 
results. Less trivial is the fact that the same effective potential does equally 
well for the ground and excited states. The consideration of the results for the 
energy-dependent model ($V_E$), where none of the above properties are a priori 
fulfilled, allows one to make more precise statements about the validity of the 
effective potential.

\subsection {Discussion} 
The present calculations with effective interactions ($V^{nr}_{eff,sc}$ and 
$V^{nr}_{eff}$) have neglected various corrections of relativistic order. In 
these cases, the binding energies  obtained are likely to be overestimated. The 
largest correction would come from the normalization factors $\frac{m}{e}$, that 
have a repulsive character. Their reduction effect tends to increase with the 
boson mass, with the angular momentum and the radial excitation. 
For the $l=0$ ground state, it is of the order of 25\% for $\mu=0.15 \, m$ and 
60\% for $\mu=0.5 \, m$. These corrections (which get partly cancelled however 
by the attractive second term neglected in Eq. (\ref{Ch})) have a size 
comparable to those obtained by Elster et al. \cite{ELST} in a schematic 
description of the deuteron or later on by Mangin-Brinet and Carbonell 
\cite{MANG} (see also ref. \cite{BILA} for the zero-mass boson case). They  are 
not small, but in the present case they should be compared to the difference 
between results obtained with $V^{nr}_{eff,sc}$ and IA$^{nr}$ which roughly 
amount to a factor 2-3 $(\mu=0.15\,m)$ and 6 $(\mu=0.50\,m)$. Curiously, 
the discrepancy with the Bethe-Salpeter results is rather small in many cases. 
In view of the corrections due to the factors $\frac{m}{e}$,  
the agreement should therefore be considered with caution. 
The effect of these factors is actually welcome since it will bring values 
obtained with $V^{nr}_{eff,sc}\; (-{\rm box})$ closer to those obtained with 
$V_E \;(-{\rm box})$, leaving some room for the contribution of Z-type diagrams 
as well as diagrams with higher order boson exchange, whose role should increase 
with the coupling. Both  have an attractive character for the field-theory model 
considered in this work. On the other hand, we would like to mention the role of 
the term $E-e-e'$ in the meson propagator, Eq. (\ref{Bb}), and the corresponding 
one, $V^{nr}_{eff,sc}$, introduced in the self-consistent  derivation of the 
effective interaction, Eq. (\ref{Cy}). The effect of these terms is especially 
important when the coupling becomes large, though the binding energy remains 
small. Properly taking into account these factors is essential to get a reliable 
estimate of the two-boson exchange contribution both in the energy-dependent and 
in the effective interaction schemes. Once this was done correctly, 
contributions dropped by one or two orders of magnitude for the second set 
of values given in Table \ref{tab2} ($\mu=0.5\,m$), bringing the results in 
an acceptable range. Contrary to the statement made in ref. \cite{MANG}, it is 
possible to roughly account in a non-relativistic approach for binding energies 
obtained in relativistic calculations even for  large couplings, provided 
one includes in the interaction corrections that have the range of multi-boson 
exchange.

Keeping in mind the above remarks about the non-relativistic calculations, it is 
instructive to look at the convergence of various approximate results to 
the ones obtained from solving the Bethe-Salpeter equation in the ladder 
approximation. For instance, the binding energies obtained with this equation 
for the $l=0$ ground state and $\mu=0.15 \, m$ are equal to  0.0059, 0.023, 
0.047, 0.076, 0.109  for $\alpha=$ 0.50, 0.75, 1.00, 1.25, 1.50 respectively
(see Table \ref{tab1}).  They can be compared to the binding energies 
calculated in the light-front approach, 0.0057, 0.022, 0.044, 0.070, 0.100 
\cite{MANG}  and with the effective interaction, $V_{eff}$,  0.0040, 0.014, 
0.026, 0.037, 0.047 \cite{AMGH}. The difference, which is dominated in both 
cases by the contribution of the box diagram, is significantly smaller in the 
former one \cite{FRED}. While this contribution is largely sufficient to 
reproduce the Bethe-Salpeter results in  the light-front approach, 
it is not for approaches using energy-dependent and effective interactions. In 
these cases, one has to incorporate also extra contributions due to Z type 
diagrams (among other ones) to account for the Bethe-Salpeter results. 

While the superiority of the light-front approach in calculating binding 
energies is obvious for massive bosons, the advantage for zero-mass bosons is 
less striking. With this respect, it is interesting to compare the 
corresponding binding energies obtained from a variational calculation 
with present ones. Expressions given by Ji and Furnstahl \cite{JI} for the $l=0$ 
and $l=1$ states allow one to derive an effective coupling in terms of $\alpha$ 
in each case, and it is found that they compare well with the ones given in 
Fig. \ref{fig6}.

For the zero-mass boson case, present results are obviously in agreement with 
those obtained in ref. \cite{SILV}. They can be compared to those given 
by Bilal and Schuck (Table 1 of ref. \cite{BILA}). From the examination of their 
approach, it is found that the memory-function results with vacuum correlations 
neglected (MFN) should be close to the ones obtained with the energy-dependent 
approach without box contribution, denoted $V_E(-{\rm box}) $. For $\alpha=1$, 
the numbers are 0.607-1 and 0.637-1 respectively, indicating that the dominant 
energy-dependent effect is accounted  for in both approaches. The box 
contribution is absent in the MFN calculation, where it appears on a different 
footing and should be considered separately. It is probably a sizeable part of 
what the memory function (MF) result is missing to reproduce the Bethe-Salpeter 
result.  This is confirmed by the comparison of this difference with our 
estimate of the box contribution. For $\alpha=1$, numbers for the lowest state 
are respectively 0.100-1 for B.S. $-$ MF and 0.088-1 for $V_E-V_E(-box)$ or 
0.121-1 for $V^{nr}_{eff,sc}-V^{nr}_{eff,sc}(-box)$. Finally, the difference 
between the MF and MFN results, which is due to vacuum correlations, should be 
comparable to what our results for $V_E$ miss to reproduce the B.S. results 
(Z-diagrams essentially). This is again verified, numbers being 0.135-1 for 
MF $-$ MFN and 0.117-1 for B.S. $-$ $V_E$  for the same case 
($\alpha=1$, $l=0$).

In all cases, the Bethe-Salpeter, the energy-dependent and the non-relativistic 
effective interaction approaches produce smaller binding energies than the 
instantaneous approximation. This would indicate that the range of validity of a 
non-relativistic approach should be larger than what is often infered 
from the results of the instantaneous approximation. How this conclusion is 
modified when going beyond the ladder approximation is discussed in the next 
section.

\section{Results with crossed-boson exchange}
We consider in this section the role of genuine crossed two-boson exchanges 
in the different approaches, including in particular those ones that have the 
same order as the correction brought about by the renormalization 
factor  $\left(1+V_1(r)\right)^{-1}$ in Eq. (\ref{Ci}) or what generalizes this 
quantity in Eq. (\ref{Cj}). We first remind the general expression for these 
contributions and subsequently look at particular cases for both
neutral and charged bosons.
 
\subsection{General case}

\begin{figure}[htb]
\begin{center}
\mbox{\psfig{file=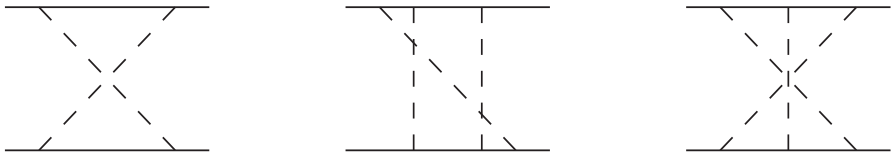,width=12cm,height=2cm}}  
\end{center}
\caption{Two- and three-crossed boson exchange contributions to the two-body 
interaction.\label{fig7}}
\end{figure}

As invoked in Sect. 2, the full kernel of the Bethe-Salpeter equation,
Eq. (\ref{Aa}), also contains diagrams of the non-ladder type. The simplest 
examples of these are shown in Fig. \ref{fig7}, involving crossed two- and 
three-boson exchanges.
The full expression of the crossed two-boson exchange is taken from ref. 
\cite{THEU}, where it was put into the form of a dispersion relation appropriate 
to be used with the Bethe-Salpeter equation, Eq. (\ref{Aa}). Among the different 
expressions given in this work, we retain here that one which is independent of 
the energy, with $u=0$:
\begin{equation}
K^{(2)}=- \frac{g^4}{8\pi^2} \int_{4\mu^2}^{\infty} \frac{dt'}{t'-t} 
\; \sqrt{ \frac{t'-4\mu^2}{t'} } \;\; 
\frac{1}{2\left( \mu^4 + m^2(t'-4\mu^2)\right)} .
\label{Ha}
\end{equation}
Provided that the binding energy is not too large, the other expressions give 
close results. 

\begin{figure}[htb]
\begin{center}
\mbox{\psfig{file=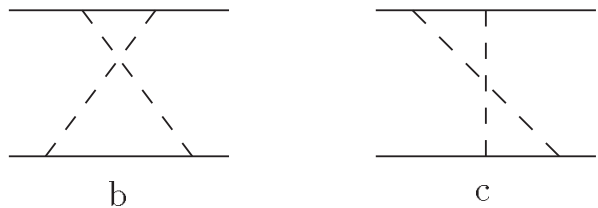,width=8cm}}  
\end{center}
\caption{Crossed two-boson exchange contributions to the two-body 
interaction in time-ordered perturbation theory. \label{fig8}}
\end{figure}

The expressions appropriate for the energy-dependent picture are obtained from 
the time-ordered diagrams shown in Fig. \ref{fig8} that have to be added to the
box diagram of Fig. \ref{fig3} a. They read:

\begin{eqnarray}
\nonumber
V^{(2b)}(\vec{p\,},\vec{p}\,') =- \frac{g^4}{2} \int \frac{d 
\vec{q}}{(2\pi)^3} \;  \frac{m}{e_q} \frac{m}{e} \;
\frac{O_{1} \, ( O_{1} \cdot O_{2}) \, O_{2}}{\omega \, \omega'} \;
\frac{m}{e'} \frac{m}{e_{q'}} \hspace{2cm} \\
\times \frac{1}{\left(\omega+e+e_q-E\right)} 
\frac{1}{\left(\omega'+e'+e_q-E\right)} 
\frac{1}{\left(\omega+\omega'+e_q+e_{q'}-E\right)},
 \label{Hb1} \\
\nonumber \rule[0em]{0em}{3em}
V^{(2c)}(\vec{p\,},\vec{p}\,') =- \frac{g^4}{2} \int \frac{d 
\vec{q}}{(2\pi)^3} \;  \frac{m}{e_q} \frac{m}{e} \;
\frac{O_{1} \, ( O_{1} \cdot O_{2}) \, O_{2}}{\omega \, \omega'} \;
\frac{m}{e'} \frac{m}{e_{q'}} \hspace{2cm} \\
\nonumber
\times \frac{1}{\left(\omega+e+e_q-E\right)} 
\frac{1}{\left(\omega+e'+e_{q'}-E\right)} \hspace{2cm} \\
\times 
\left\{ \frac{1}{\left(\omega+\omega'+e+e'-E\right)} +
\frac{1}{\left(\omega+\omega'+e_q+e_{q'}-E\right)} \right\}.
 \label{Hb2} 
\end{eqnarray}

In addition to the factors defined after Eq. (\ref{Bf}), we have used here 
$e_{q'}=\sqrt{m^2+\vec{q}\,'\,^2}$ with 
$\vec{q}\,'=\vec{q}-\left( \vec{p}+\vec{p}\,' \right)$.
Z-type diagrams have a higher order in $\frac{1}{m}$ and are expected 
to be smaller. They are not incorporated in Eqs. (\ref{Hb1}, \ref{Hb2}).
By retaining terms of the lowest order, $\left(\frac{1}{m}\right)^0$, one gets 
the expression of the crossed two-boson exchange that we will include in 
$V_{eff,sc}^{nr}$ and $V_{eff}^{nr}$. It can be checked that it is 
equal to the contribution obtained with the dispersion relation, 
Eq. (\ref{Ha}), in the same limit. 

For a charged boson, the crossed-box contribution depends on the 
isospin of the state under consideration. We will consider the 
case of a state with $T=1$, where it provides further attraction. 
The effective interaction then reads:
\begin{equation}
V^{nr}_{eff}=V_0 \; \frac{ 1+\bar{V}^{(2a)}+\kappa \, 
(\bar{V}^{(2b)}+\bar{V}^{(2c)})}
{1+\bar{V}^{(2a)}+\bar{V}^{(2b)}+\bar{V}^{(2c)}},
\label{Hc}
\end{equation}
where the barred potentials  are the same as the ones defined in Eqs. 
(\ref{Bf}, \ref{Hb1}, \ref{Hb2}) but without the isospin operators $O_i$.
The factor $\kappa$ is the result of the isospin algebra. For a charged 
boson, $\kappa=5$, whereas for a neutral boson, $\kappa=1$, in which case 
the expression greatly simplifies, leaving $V_0$ as an effective interaction. 
Notice that, up to order $\left(\frac{1}{m}\right)^0$, this cancellation is 
expected to hold at all orders in the coupling $g^2$, i.e. even including 
multi-boson exchange diagrams \cite{AMGH}. The effective interaction is thus 
also given by $V_0$ in this case.

\subsection{Massive spinless bosons (without and with ``charge'')}

\begin{table}[p]
\caption{\label{tab4}
Ground-state binding energies calculated for massive, scalar bosons without 
and with charge (in units of the constituent mass m), including the box- and 
the crossed-box contribution. Results for the charged-boson case correspond 
to a state with total isospin equal to 1. As in the previous section, five 
cases have been considered: the Bethe-Salpeter equation (B.S.), an 
energy-dependent model ($V_E$), an effective interaction ($V^{nr}_{eff,sc}$), 
an approximate version of the latter one ($V^{nr}_{eff}$) and, slightly apart, 
the so-called instantaneous approximation(IA$^{nr}$). }
\begin{tabular}{cccccccc}
\hline 
\rule{0pt}{3.3ex}
  $\mu/m$ &$ \alpha$   & ${\rm B.S.}$ & $V_E$ & $V^{nr}_{eff,sc}$ & 
  $V^{nr}_{eff} $   & \hspace*{0.5cm}& ${\rm IA}^{nr} $ \\ [1.ex]
\hline 
 $ {\rm neutral~case }$ &        &        &       &       & &         \\ [1.ex]
 $     $&$    0.50    $ & 0.013  & 0.0076 & 0.0096& 0.013 & & 0.013   \\ [0.ex]
 $     $&$    0.75    $ & 0.062  & 0.027  & 0.035 & 0.056 & & 0.056   \\ [0.ex] 
 $ .15 $&$    1.00    $ & 0.158  & 0.054  & 0.071 & 0.129 & & 0.129   \\ [0.ex]
 $     $&$    1.25    $ & 0.324  & 0.083  & 0.114 & 0.233 & & 0.233   \\ [0.ex]
 $     $&$    1.50    $ & 0.593  & 0.115  & 0.161 & 0.367 & & 0.367   \\ [1.ex]
 $     $&$    1.50    $ & 0.077  & 0.014  & 0.034 & 0.093 & & 0.093   \\ [0.ex]
 $     $&$    1.75    $ & 0.162  & 0.029  & 0.064 & 0.180 & & 0.180   \\ [0.ex]
 $ .50 $&$    2.00    $ & 0.290  & 0.047  & 0.101 & 0.296 & & 0.296   \\ [0.ex]
 $     $&$    2.25    $ & 0.479  & 0.068  & 0.143 & 0.442 & & 0.442   \\ [0.ex]
 $     $&$    2.50    $ & 0.761  & 0.092  & 0.189 & 0.619 & & 0.619   \\ [1.ex]
 ${\rm charged~case } $ &        &        &               & &         \\ [1.ex]
 $     $&$    0.30    $ & 0.0062 & 0.0015 & 0.0024& 0.0095& & 0.00046 \\ [0.ex] 
 $     $&$    0.40    $ & 0.037  & 0.0083 & 0.012 & 0.053 & & 0.0046  \\ [0.ex] 
 $ .15 $&$    0.50    $ & 0.107  & 0.020  & 0.028 & 0.152 & & 0.013   \\ [0.ex]
 $     $&$    0.60    $ & 0.229  & 0.034  & 0.049 & 0.322 & & 0.027   \\ [0.ex] 
 $     $&$    0.70    $ & 0.425  & 0.050  & 0.073 & 0.573 & & 0.045   \\ [1.ex]
 $     $&$    0.90    $ & 0.092  & 0.0007 & 0.011 & 0.336 & & 0.0007  \\ [0.ex] 
 $     $&$    1.00    $ & 0.183  & 0.0041 & 0.023 & 0.59  & & 0.0052  \\ [0.ex]
  \raisebox{1.5ex}[0pt]{$.50$}
        &$    1.10    $ & 0.319  & 0.0098 & 0.040 & 0.93  & & 0.0138  \\ [0.ex]
 $     $&$    1.20    $ & 0.517  & 0.0170 & 0.061 & 1.36  & & 0.0267  \\ [1.ex]
 
\hline
\end{tabular}
\end{table}

The results for the massive boson case are given in Table \ref{tab4}. 
As in the ladder approximation, results for the energy-dependent and the 
non-relativistic self-consistent effective interaction approaches are 
relatively close to each other in the neutral-boson model. 
The discrepancy, which tends to increase with the coupling and the boson mass, 
can be understood as being primarily due to the omission of factors 
$\sqrt{\frac{m}{e}}$ in the non-relativistic calculation. Taking this into 
account, it can be considered that the effective self-consistent interaction, 
$V^{nr}_{eff,sc}$, represents the physics underlying the energy-dependent 
interaction, $V_{E}$, which it was derived from. However, contrary to the 
ladder approximation (Table \ref{tab1}), the corresponding results 
significantly depart from those obtained with the Bethe-Salpeter 
equation or the (non self-consistent) effective interaction, $V^{nr}_{eff}$, 
which both are close (or identical) to the results obtained in the 
instantaneous approximation. It is reminded that the exact equality of the two 
last columns simply results from the identity, $V_{eff}^{nr}= V_0$, which 
holds for this particular case, see Eq. (\ref{Hc}) for $\kappa=1$. 
We notice here that the discrepancy between the two sets of results tends 
to increase with the coupling or the boson mass. This feature has a general 
character in the domain and points to the larger role of higher-order 
contributions that should show up simultaneously (Z-diagrams or multi-boson 
exchange). 

Discarding results in the instantaneous approximation, those obtained in the 
charged-boson case show qualitative features similar to the neutral-boson 
case. Results rather go by pair, $V^{nr}_{eff,sc}$ and $V_{E}$ on one side, 
B.S. and $V^{nr}_{eff}$ on the other. Relative differencies however are 
larger. On the one hand, most results are larger than those obtained with 
the instantaneous approximation. While the latter one was apparently  
supported by the results for the neutral-boson case, the charged-boson 
ones show that this approximation is a poor one. On the other 
hand, a feature evidenced by the examination of Table \ref{tab4} is the 
smaller binding energy obtained with the Bethe-Salpeter equation as compared 
to what is  obtained with the effective interaction $V_{eff}^{nr}$. 
This suggests that the closeness of the corresponding results in the neutral 
boson case is partly misleading. As mentioned for the ladder approximation, 
the effective interaction misses attractive contributions due to Z-type 
diagrams while the Bethe-Salpeter approach involves some renormalization 
which reduces the attraction. These effects may be  roughly equal in the 
neutral-boson case but could differ substantially in the charged-boson one.

\subsection{Discussion}

By incorporating the contribution of crossed diagrams, we expected 
the diffe\-rent approaches, B.S., $V_{E}$, $V_{eff,sc}^{nr}$ and $V_{eff}^{nr}$, 
to produce a pattern similar to the one evidenced by these models in the ladder 
approximation. As shown in the previous subsection, this is not the case. An 
important remark is that, contrary to the ladder approximation, we miss here 
reliable benchmark results such as the B.S. ones there. The results for 
the B.S. approach given in Table \ref{tab4} rely on dispersion relations for 
calculating the crossed-boson exchange contribution \cite{THEU}. These ones are
well defined within a certain procedure to deal with off-shell effects but this 
procedure itself is not unique. This prevents one from making a 
meaningful comparison of the different approaches without entering into a 
detailed examination. It is only in the case where it has a perturbative 
character that this examination could be skipped. A different benchmark is 
that one provided by $V_{eff}^{nr}(={\rm IA}^{nr})$ in the neutral boson case, 
which can be seen to represent the sum of all contributions at the lowest 
order $(\frac{1}{m})^0$ \cite{AMGH}. A last benchmark is given by the results 
obtained by Nieuwenhuis and Tjon using the Feynman-Schwinger representation 
\cite{NIEU}. They incorporate implicitly all contributions of crossed-boson 
exchanges. In comparison, the crossed-box exchange contribution calculated 
here with the B.S. equation only explains about half of the total contribution 
calculated by these authors.

Since $V_{eff}^{nr}$ represents at the same time the sum of all contributions 
(box and crossed box, see ref. \cite{AMGH}) and the contributions with the 
range of two-boson exchange (see Eq. (\ref{Hc}) for $\kappa=1$), it could 
be infered that the convergence in terms of the number of exchanged bosons 
is quickly achieved. This conclusion has an ambiguous character however since 
accounting for all contributions with the range of three boson exchange also 
allows one to recover the expected total sum which is nothing but the IA$^{nr}$ 
result for $\kappa=1$. A better insight on the role of three, four... 
boson exchange is provided by the examination of the difference 
between results obtained with $V_{eff}^{nr}$  and with the 
self-consistent potential, $V_{eff,sc}^{nr}$.  This likely 
represents an underestimate of the effect however. These extra contributions 
will reduce the part due to two-boson exchange in the effective interaction. 
The comparison of results obtained with $V_{eff}^{nr}$ at the order of 
two-boson exchange, and those obtained with $V_{E}$ provides 
an estimate on the role of various ingredients entering the calculation: 
factors  $\sqrt{\frac{m}{e}}$ and ($E-e-e'$)-corrections  in the meson 
propagators. The large effect prevents one from considering the results 
obtained with $V_{eff}^{nr}$ as really representative of a complete 
calculation. This observation is especially important when the comparison 
is made  with results obtained with the B.S. approach, which they 
roughly agree with, or with those of Nieuwenhuis and Tjon.
 
At first sight, results obtained with the B.S. approach seem reasonable if 
they are compared to the FSR results. They leave some room 
for contributions of three- or multi-boson exchange, which should be roughly 
equal to the two-boson exchange contribution for the largest couplings for 
which a comparison is possible. The large discrepancy between our results for 
the B.S. and $V_{E}$ approaches suggests that the situation is not so simple 
however. This discrepancy, which should be due in first approximation to 
Z-type diagrams, is much larger than in the ladder approximation, as shown 
in Table \ref{tab1}. This may be partly due to the possible proximity of 
a situation where the binding energy would rapidly increase while the 
coupling approaches a critical value. More probably, it seems that the 
dispersion-relation approach does not take fully into account the 
renormalization effects that $V_{E}$ and $V_{eff,sc}^{nr}$ implicitly 
incorporate. This situation has some similarity with that one encountered 
for $V_{eff}^{nr}$, which, for the crossed two-boson exchange, turns 
out to provide results identical to the instantaneous approximation ones. If 
correct, the above interpretation would have consequences for approaches based 
on dispersion relations for calculating crossed-box diagrams. In the simplest 
case, beside the full three crossed-boson exchange contribution, there should 
be destructive contributions of the same range, corresponding to the topology 
(crossed-box $\otimes$ single-boson) exchange. 

While the present work was completed, we were informed about different results 
accounting for the full crossed-boson exchange diagram \cite{TJON3}. 
These ones are close to those obtained with $V_{E}$ or $V_{eff,sc}^{nr}$, 
possible discrepancies in the first case being likely due to the contribution 
of Z-type diagrams omitted here. They tend to support the idea that calculating 
the crossed-box with dispersion relations is not the best way to provide an 
estimate of its contribution as soon as the coupling increases. Despite the 
fact that  the obtained values are far from  reproducing the FSR 
results for large couplings, they could still be an overestimate.

\section{Conclusion}
In this work we investigated different approaches based on the Bethe-Salpeter 
equation, in particular a 3-dimensional equation with an energy-dependent 
interaction and a non-relativistic equation with an energy-independent 
effective interaction. This has been done for the ladder approximation 
as well as the more complete approach involving crossed-boson exchange 
with the idea that approximation schemes should not be limited to one 
of these cases. We especially compared the binding energies for the 
lowest (normal) states for the case of distinguishable scalar particles 
with masses for the exchanged boson varying from 0 to $0.5\,m$. This 
work completes and improves upon another one where the emphasis was 
rather put on the wave functions obtained from the last two approaches, 
with and without energy-dependent interactions.   Up to departures of 
relativistic order, we found a strong continuity between the results 
from different approaches in the ladder approximation.  The inclusion 
of crossed-diagram contributions points to a strong sensitivity to the 
way the problem is approached.

The reasons why we are able to account in a non-relativistic approach for 
results of the Bethe-Salpeter equation in the ladder approximation deserve 
some explanation. The difference between the results obtained with the 
Bethe-Salpeter equation and those using the instantaneous single-boson exchange 
approximation in a non-relativistic approach is not due to relativity, as 
sometimes advocated, but to the fact that the latter misses the field-theory 
character underlying the former. Since a field-theory can be dealt with in a 
non-relativistic scheme, it should be no surprise that one is able to get rid 
of the above difference. The so-called instantaneous approximation, often 
employed to provide some non-relativistic interaction, simply represents 
a too naive view of what the interaction in a non-relativistic  picture 
should be, missing many-body effects in its determination. Contrary to 
a common belief, the dynamics involving the relative time variable can be 
accounted for partly, if not totally, in a three-dimensional approach, 
in a way which is  not Lorentz-covariant however. The main effect in 
the small coupling limit, which implies corrections with the range of 
multi-boson exchange, is a renormalization of the so-called instantaneous 
interaction by  the probability of having two constituents only. Since in the 
ladder-approximation, the part with in-flight bosons plays no role, the 
force turns out to be effectively weakened.  This is one of the most 
intuitive effects one can think of. The intensity of the force being 
smaller, it becomes easier for a self-consistent, non-relativistic 
approach to account for effects obtained in a relativistic one. The 
improvement we introduced in calculating the effective interaction is quite 
significant. While it removes partly the agreement with more elaborate 
calculations achieved in some cases, we believe that the effect is 
fully relevant. This points to ingredients missing in the effective 
interaction, especially those in relation with non-locality effects, 
which are obviously more difficult to incorporate. 

We also looked at the contribution due to crossed two-(massive) boson 
exchange. This exchange represents an essential improvement upon the 
ladder approximation, which offers interest with certain respects but 
remains academic with other ones. The effects are generally quite large, 
especially in the case of charged-boson exchange where one should be seriously 
concerned by higher order crossed-boson exchange contributions. In any case, 
this casts serious doubts on the possibility to reliably 
describe physical systems in the strong-interaction regime by only 
retaining one-boson exchange contributions. This could be however partly 
remedied by fitting parameters of a certain model, when possible, as 
it is done for the NN interaction models. In view of these large effects, 
the fact that, in some cases, we recover more or less the results of the 
so-called instantaneous approximation for the neutral-boson exchange  
should be considered with a lot of caution. This result is obtained 
because the contribution of the crossed two-boson exchange, 
most often neglected, has been included consistently with the box 
two-boson exchange contribution. Such a result may have a more general 
character and be simply due to the fact that, for neutral, spinless-boson 
exchange, the interaction, supposed to be  given by the instantaneous 
approximation, is the same for components with an undetermined number of 
accompanying bosons and, therefore, transparent to the presence of bosons 
in flight. By no means, one can draw an argument from this result in 
favor of the validity of the one-boson exchange contribution in the 
non-relativistic approach. 

This conclusion is supported by another feature of our results. The binding 
ener\-gies calculated with the Bethe-Salpeter equation and the effective 
interaction are significantly smaller than the completely non-perturbative 
ones obtained by Nieuwenhuis and Tjon \cite{NIEU} (a factor of 2 
at $\alpha \simeq 0.9$), indicating the existence of further sizeable 
corrections of higher order \cite{THEU}. Most probably, the closeness 
of our results for the Bethe-Salpeter equation and the effective 
interaction, which was obtained in a few cases, is partly 
accidental, as already mentioned above. Those with the effective 
interaction miss contributions of relativistic order, like Z-type diagrams, 
while the Bethe-Salpeter equation, where these ones are included, 
does not take into account contributions with a higher number of exchanged 
bosons. The accidental character of the closeness of these different results 
is further supported by the examination of those we obtained for the 
charged boson case.

We presented results for zero-mass bosons only in the ladder approximation. 
Some results were obtained for the crossed-boson exchange diagram but 
they did not evidence any striking difference with the $\mu=0.15\,m$ case. 
These results need to be more elaborated, especially in view of possible 
applications to the photon or gluon case. The cancellation between 
the renormalization of the instantaneous approximation interaction and the 
effect of the crossed-boson exchange, which is expected in the massive 
neutral-boson case at order $\left(\frac{1}{m}\right)^0$, and has been partly 
obtained in our calculations, is more difficult to be achieved in 
the zero-mass case. For small couplings, the corrections to the binding 
energies that are dominated by the term $\frac{\alpha^2}{4n^2}$ are 
of the order $\alpha^3\, \log (c\alpha)$ \cite{FELD}, while the total correction is 
expected to be of order $\alpha^4$.  The reason is that solving a certain
integral equation allows one to sum up an infinite set of contributions 
with increasing order, but also an increasingly divergent character in the 
infra-red limit. This prevents one from reproducing numerically the above 
expected cancellation without considering an infinite set of selected 
crossed-boson exchanges, having similar properties. From present results 
and further studies performed separately, it seems that the convergence 
to the full result is likely to be slow for couplings of the order 
$\alpha \simeq 1$.
 
All studies presented here have been performed in a model involving scalar 
particles only. This greatly simplifies the discussions by providing a better 
insight on the origin of various corrections (renormalization of the 
instantaneous approximation interaction, relativistic corrections,....). 
Despite the fact that the model is not an especially realistic one, we 
nevertheless believe that our conclusions have some general character. 
The most important one is probably the very limited validity range of 
the instantaneous one-boson exchange approximation which, strictly 
speaking, is valid only for small energy transfers and therefore  only 
makes sense for the Born amplitude in the non-relativistic regime. 
Its apparent success in QED or in hadronic physics is largely misleading. 
It rather points to large contributions from multi-boson exchanges. In 
the case of QED, they are required to recover the Coulomb potential as 
is well known, whereas in hadronic physics, they provide a more microscopic 
description of the forces, most often simulated by instantaneous 
exchanges of single bosons with fitted parameters.

{\bf Acknowledgements}
We are indebted to E. Moulin and N. Ploquin, whose work 
on non-locality effects provided useful insight on their role in 
the context of the present work. We are also grateful to D. Phillips for 
providing results concerning the contribution of the crossed-box diagram.

\begin{appendix}
\section{The box diagram \label{appA}}

We give here the analytic result for the box diagram of Fig. \ref{fig3} 
a, in the case where the factors $\frac{m}{e_q}$ and terms like $E-e-e'$ 
in the denominator of Eq. (\ref{Bf}) are neglected. It is not what we 
actually used in the calculations presented in this work 
but it turned out to be quite useful 
for numerical tests. The (energy-independent) expression for 
$V^{(2a)}(\vec{p\,},\vec{p}\,')$ then takes the form

\begin{equation}
V^{(2a)}(\vec{p\,},\vec{p}\,') =- \frac{1}{2} \int \frac{d 
\vec{q}}{(2\pi)^3} \;   \frac{m}{e}  \left(  \frac{g^{4}  \; 
(O_{1} \, O_{2}) \, 
(O_{1} \, O_{2})}{\omega \, \omega'\; (\omega +\omega') } 
 \;\; \frac{1}{\omega \, \omega' }\right)\frac{m}{e'},
 \label{box0} 
\end{equation}
with the definitions 
$\omega=\sqrt{\mu^{2}+(\vec{p}-\vec{q}\,)^{2}}$, 
$\omega'=\sqrt{\mu^{2}+(\vec{p}\,'-\vec{q}\,)^{2}}$,
$e=\sqrt{m^{2}+\vec{p}^{\,2}}$ and $e'=\sqrt{m^{2}+\vec{p}\,'^{\,2}}$.
If one
defines the integral $ I$ by
\begin{equation}
\label{box1}
V^{(2a)}(\vec{p\,},\vec{p}\,') =: -\frac{1}{2} \frac{m^2}{e e'}
\frac{g^4}{(2\pi)^{3}} (O_{1} \, O_{2}) \, (O_{1} \, O_{2}) \,  I,
\end{equation}
then it is found that
\begin{eqnarray}
\label{box2}
 I &=& \int \frac{d\vec{q}}{\left[ \mu^2 +(\vec{p}-\vec{q})^2\right] 
\left[ \mu^2 +(\vec{p}\,'-\vec{q})^2\right] 
\left[ \sqrt{\mu^2 +(\vec{p}-\vec{q})^2} + 
\sqrt{\mu^2 +(\vec{p}\,'-\vec{q})^2}\right] } \nonumber \\
 &=& \frac{2\pi}{k}\int_0^{\infty} \frac{q\;dq}{\left(
 \mu^2+q^2\right)^{\frac{3}{2}}}
 \log \frac{\sqrt{\mu^2 +(k+q)^2}
 \left[ \sqrt{\mu^2+q^2} + \sqrt{\mu^2 +(k-q)^2} \right]}
 {\sqrt{\mu^2 +(k-q)^2}
\left[ \sqrt{\mu^2+q^2} + \sqrt{\mu^2 +(k+q)^2} \right]}, \nonumber \\
\end{eqnarray}
where $k=|\vec{k}| = |\vec{p} - \vec{p}\,'|$. From this form
one derives the following analytical properties of the integral $ I$:
\begin{eqnarray}
\nonumber
 I = \frac{2\pi}{3\mu^2} \hspace{1cm} \;\;\; & {\rm if} & \;\; \vec{k}=0 , \\
\label{box4}
 I = \frac{2\pi}{k^2} \;  4\log 2 \;\;\; & {\rm if} & \;\; \mu=0 , \\
\nonumber
 I \rightarrow \frac{2\pi}{k^2} \; 4\log2 \;\;\; & {\rm for} & \;\; k 
\rightarrow \infty .
\end{eqnarray}
Integrating by parts, one finds the following expression for $ I$:
\begin{eqnarray}
\label{box6} \nonumber
 I &=& \frac{2\pi}{k\sqrt{k^2+4\mu^2}} \left\{ 
\frac{2}{\sqrt{1-c}} \log \frac{1+\sqrt{1-c}}{1-\sqrt{1-c}} \right.\\
&& \hspace{3cm}\left. -4\sqrt{\frac{c}{1-c}} \arctan \sqrt{\frac{1-c}{c}} + 2\log c
\right\},
\end{eqnarray}
where the factor $c$ is given by 
\begin{equation}
\label{box7}
c=\frac{\sqrt{k^2+4\mu^2}-k}{\sqrt{k^2+4\mu^2}+k}.
\end{equation}
A careful analysis reveals that the expression of Eq. (\ref{box6}) satisfies the
conditions of Eqs. (\ref{box4}), 
because of a cancellation of divergencies in single terms.
\end{appendix}


\end{document}